\documentclass[10pt]{article}

\usepackage{multirow}
\usepackage{booktabs}
\usepackage{tabularx}
\usepackage{ltablex}
\usepackage{array}

\usepackage{amssymb,amsmath,amsthm}
\usepackage{nicefrac,bbm}
\theoremstyle{definition}   
\usepackage{apptools,mathtools,chngcntr}

\usepackage{color,xcolor}
\usepackage{tikz,pgfplots}
\usepgfplotslibrary{colormaps,fillbetween,groupplots}
\usepackage{pgfplotstable}
\usepackage{subcaption}
\usepackage[font=small]{caption}
\usepackage{wrapfig}
\pgfplotsset{compat=newest}
\usetikzlibrary{patterns,shapes,arrows.meta,patterns.meta}
\tikzset{>=latex} 
\usepackage{pgfgantt}

\usepackage[maxbibnames=50, maxcitenames=2, uniquelist=false,
    natbib=true, style=authoryear-comp,
    isbn=false, url=false,giveninits=true,dashed=false]{biblatex}   
\bibliography{bibliography}
\AtBeginBibliography{} 

\usepackage[utf8]{inputenc}
\usepackage[T1]{fontenc}
\usepackage[a4paper,left=25mm,right=25mm,top=25mm,bottom=25mm]{geometry}
\usepackage[hidelinks]{hyperref}
\usepackage{pdfpages}
\usepackage[capitalise]{cleveref}
\crefname{section}{Section}{Sections}
\crefname{appendix}{Appendix}{Appendices}
\crefname{table}{Table}{Tables}
\usepackage{titlesec, blindtext}
\usepackage{marginnote}

\newtheorem{definition}{Definition}[section]
\newtheorem{remark}[definition]{Remark}

\usepackage[affil-it]{authblk}
\usepackage{enumitem}

\newcommand{\tW}{\widetilde{W}}

\newcommand{\ba}{\boldsymbol{a}}
\newcommand{\bb}{\boldsymbol{b}}

\newcommand{\be}{\boldsymbol{e}}

\newcommand{\bn}{\boldsymbol{n}}

\newcommand{\bw}{\boldsymbol{w}}
\newcommand{\bx}{\boldsymbol{x}}

\newcommand{\bA}{\boldsymbol{A}}
\newcommand{\bB}{\boldsymbol{B}}
\newcommand{\bC}{\boldsymbol{C}}

\newcommand{\bF}{\boldsymbol{F}}
\newcommand{\bG}{\boldsymbol{G}}
\newcommand{\bH}{\boldsymbol{H}}
\newcommand{\bI}{\boldsymbol{I}}

\newcommand{\bM}{\boldsymbol{M}}

\newcommand{\bP}{\boldsymbol{P}}
\newcommand{\bQ}{\boldsymbol{Q}}

\newcommand{\bW}{\boldsymbol{W}}

\newcommand{\bphi}{\boldsymbol{\phi}}

\newcommand{\bxi}{\boldsymbol{\xi}}

\newcommand{\cD}{{\mathcal{D}}}

\newcommand{\cG}{{\mathcal{G}}}

\newcommand{\cI}{{\mathcal{I}}}

\newcommand{\cK}{{\mathcal{K}}}
\newcommand{\cL}{{\mathcal{L}}}

\newcommand{\cP}{{\mathcal{P}}}

\newcommand{\cS}{{\mathcal{S}}}

\newcommand{\bbA}{{\mathbb{A}}}

\newcommand{\bbH}{{\mathbb{H}}}
\newcommand{\bbI}{{\mathbb{I}}}

\newcommand{\bbN}{{\mathbb{N}}}

\newcommand{\bbR}{{\mathbb{R}}}

\newcommand{\bnull}{\boldsymbol{0}}

\newcommand{\rone}{\ba\otimes\bb}
\newcommand{\roneB}{(\ba\otimes\bb)}

\newcommand{\bcI}{\boldsymbol{\mathcal{I}}}

\newcommand{\bcP}{\boldsymbol{\mathcal{P}}}

\newcommand{\SP}{\mathcal{SP}}

\newcommand{\GL}{\text{GL}}

\newcommand{\SO}{\text{SO}}
\newcommand{\SYM}{\text{SYM}}

\newcommand{\tr}{\operatorname{tr}}

\newcommand{\cof}{\operatorname{cof}}

\newcommand{\norm}[1]{\left\lVert#1\right\rVert}
\newcommand{\normm}[1]{\big\lVert#1\big\rVert}

\newcommand{\Cross}{\mathbin{\tikz [x=1.4ex,y=1.4ex,line width=.25ex] \draw (0.1,0.1) -- (0.9,0.9) (0.1,0.9) -- (0.9,0.1);}}

\newcommand{\bMti}{\bM^{\text{ti}}}

\newcommand{\IoneISO}{I_1^{\text{iso}}}
\newcommand{\ItwoISO}{I_2^{\text{iso}}}
\newcommand{\IthreeISO}{I_3^{\text{iso}}}

\newcommand{\IoneTI}{I_1^{\text{ti}}}
\newcommand{\ItwoTI}{I_2^{\text{ti}}}

\definecolor{CPSgreen}{RGB}{22,164,138}
\definecolor{CPSlightblue}{RGB}{104,143,198}
\definecolor{CPSdarkblue}{RGB}{67,83,132}
\definecolor{CPSgrey}{RGB}{204, 204, 204}
\definecolor{CPSorange}{RGB}{246,163,21}
\definecolor{CPSred}{RGB}{194,76,76}
\definecolor{CPSdarkgrey}{RGB}{90, 90, 90}

\title{
On limitations of polyconvexity
\vspace{1ex}
}

\author[1,*]{Dominik~K.~Klein}
\author[2]{Rogelio~Ortigosa}
\author[3]{Heinrich~T.~Roth}
\author[3]{Karl~A.~Kalina}
\author[2]{\\Jes\'us~Mart\'inez-Frutos}
\author[3]{Markus~K\"astner}
\author[1]{Oliver~Weeger}
\affil[1]{\footnotesize Cyber-Physical Simulation, 
Department of Mechanical Engineering, TU Darmstadt, 64293 Darmstadt, Germany}
\affil[2]{\footnotesize Multiphysics Simulation and Optimization, TU Cartagena, Campus~Muralla~del~Mar, 30202, Cartagena (Murcia), Spain}
\affil[3]{\footnotesize Institute of Solid Mechanics, TU Dresden, 01062 Dresden, Germany}
\affil[*]{\footnotesize Corresponding author, email: klein@cps.tu-darmstadt.de}

\date{May 29, 2026}
 
\begin{document}

\maketitle 

 \begin{center}
     \textit{Dedicated to all advocates and critics of polyconvexity.}
     \vspace*{0.6cm}
 \end{center}

\par\noindent\rule{\textwidth}{0.4pt}
\begin{abstract}

Polyconvex constitutive modeling is attractive as it guarantees stability of numerical simulations and can improve the generalization behavior of material models. However, in certain applications, polyconvex formulations perform poorly in reproducing the underlying ground truth material response, which can effectively preclude their practical use. In this work, we address this issue and investigate the limitations of polyconvex constitutive modeling. The \textbf{main contributions} of this paper are as follows: \textbf{(1)} We analyze the theoretical reasons why polyconvexity may, in some cases, impose overly restrictive constraints that limit the achievable accuracy of constitutive models. Thereby, we provide analytical ellipticity guarantees for two non-polyconvex Mooney-Rivlin type potentials. \textbf{(2)} We investigate the practical limitations of polyconvex physics-augmented neural network constitutive models using two representative formulations: models using structural tensor-based invariants and models using signed singular values. Their performance is evaluated on datasets obtained from homogenized microstructured materials, and their predictive capabilities are assessed in finite element simulations. \textbf{(3)} Overall, we provide an overview of benefits, limitations, and mitigation strategies of polyconvex constitutive modeling.

\end{abstract}
\vspace*{2ex}
\par\noindent\rule{\textwidth}{0.4pt}

\section{Introduction}\label{chap:intro}

In constitutive modeling, ``[truth] is much too complicated to allow for anything but approximation'' \parencite{neumann1947}, and we rely on {assumptions} in the form of {constitutive conditions} which the models should fulfill \parencite[Sec.~5.10]{ciarlet1988}. 
Some of these constitutive conditions arise from physical reasoning, and their inclusion is essential, for instance, ensuring thermodynamic consistency.
Other conditions are mostly mathematically motivated. Their inclusion is optional and should be evaluated based on whether they benefit or limit the constitutive model performance for some specific application. One such mathematically motivated condition is {polyconvexity}. While polyconvexity has proven a useful constitutive condition in many applications, in other scenarios, polyconvex constitutive models are overly restrictive and show a poor performance. In this work, we explore the limitations of polyconvex constitutive modeling.


\subsection{An engineering perspective on polyconvexity}\label{sec:eng_perspective}

In finite elasticity theory, convexity of the potential $W$ in the deformation gradient $\bF$ alone would be overly restrictive and incompatible with a mechanically reasonable material behavior. Specifically, this would be incompatible with growth and objectivity conditions, and for strict convexity, it would fail to capture relevant phenomena like buckling \parencite{ciarlet1988}. In the seminal work by \textcite{Ball1976,Ball1977}, the \textbf{polyconvexity} condition was introduced. Polyconvex potentials are convex in an extended set of arguments, including the deformation gradient $\bF$, its cofactor $\bH=\cof\bF$, and its determinant $J=\det\bF$, making this convexity condition compatible with considerations on a mechanically reasonable material behavior. When the hyperelastic potential is polyconvex and an additional coercivity condition is fulfilled, the existence of minimizers of the underlying variational functionals of finite elasticity theory is guaranteed. Furthermore, polyconvexity implies \textbf{ellipticity} (or rank-one convexity) of hyperelastic potentials \parencite{Zee1983,neff2015}. Overall, the following implications hold:\footnote{Note that implications between constitutive conditions often require sufficient smoothness of the hyperelastic potential, which we assume throughout this work \parencite{neff2015}.}
\vspace{-0.1cm}
\begin{figure}[h!]
\centering
\includegraphics[width=0.58\textwidth]{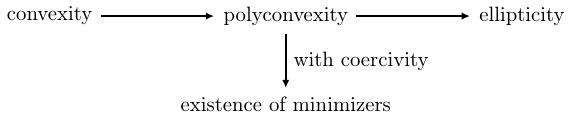}
\end{figure}
\vspace{-0.25cm}

\noindent
Note that the reversed relations do not hold in general \parencite{neff2015}. The coercivity conditions required for existence theorems impose assumptions on the material behavior which lie far outside a practically relevant deformation range, raising questions about their relevance in engineering applications \parencite{klein2022a}. Besides that, polyconvexity has no immediate physical meaning \parencite{Ball1977}, and polyconvex constitutive models can produce stress predictions that contradict mathematical considerations for idealized elastic materials \parencite{WOLLNER2026106465}. Ellipticity, in turn, ensures material stability \parencite{Schroeder_Neff_Balzani_2005}, meaning a stable and robust behavior when applying the constitutive model in numerical applications such as the finite element method. This makes ellipticity a very desirable property of the hyperelastic potential. From a physical perspective, ellipticity guarantees the existence of real-valued wave speeds for solutions of the governing equations of finite elasticity theory \parencite{Zee1983}. Without polyconvexity, the ellipticity of a hyperelastic potential is practically impossible to fulfill by construction. Thus, we employ polyconvexity as it implies ellipticity, rather than for its significance in existence theorems. Moreover, polyconvexity serves as a strong inductive bias that can improve the generalization behavior of the constitutive model, particularly for extrapolation away from the calibration data \parencite{klein2025,kalina2024a}. Notably, the polyconvexity condition has inspired a variety of numerical methods~\parencite{schroeder2011,betsch2018,bonet2015,franke2023}, and was also extended to electro-magneto-mechanical material behavior~\parencite{gil2016}, where it is again linked to existence theorems~\parencite{silhavy2018} and furthermore ensures multi-physical ellipticity~\parencite{Ortigosa_Gil_2016_hyperbol}.

While we only model elliptic material behavior in this work, it is important to note that \textbf{non-ellipticity} can be mechanically reasonable as well. Examples include the modeling of damage \parencite{KOHLER2023105199}, the homogenized behavior of microstructured materials \parencite{Rudykh_deBotton_2012}, or plasticity \parencite{Carstensen2002,bartels2006}, all of which can cause a loss of ellipticity in the constitutive model. Despite this mechanical plausibility, using non-elliptic models in numerical simulations can lead to unfavorable effects, such as mesh dependence and convergence issues of the solver \parencite{Balzani_Ortiz_2012}. To ensure stable and accurate computations in such cases, additional methods must be applied, e.g., convex envelopes of the originally non-convex formulation \parencite{balzani2024,BARTELS20045143}.
In contrast, an elliptic constitutive model automatically guarantees stable and robust behavior when used in numerical simulation methods. Thus, in the formulation of constitutive models, one should differentiate between ellipticity and a loss thereof. Moreover, for elliptic material behavior, we would like polyconvex constitutive models to perfectly represent the ground truth material behavior. In contrast, for non-elliptic material behavior, a perfect representation by a polyconvex constitutive model can be ruled out a priori.


\subsection{Polyconvex constitutive modeling and its benefits}

For almost three decades after its initial conception, polyconvex constitutive modeling was practically limited to isotropic material behavior, with models formulated in terms of the main invariants of the right Cauchy-Green tensor $\bC$ or in terms of principal stretches \parencite{ciarlet1988,Ball1976}. In the landmark work of \textcite{Schroeder2003}, for the first time, a polyconvex constitutive model applicable for anisotropic material behavior was proposed, followed by \textcite{Hartmann2003}, \textcite{Itskov2004}, \textcite{Kambouchev2007}, \textcite{Ehret2007}, and \textcite{Schroeder2008}. These approaches are based on polyconvex invariants of the right Cauchy-Green tensor $\bC$ and tuples of structural tensors. By that, the models fulfill objectivity and material symmetry a priori. Then, by a suitable construction of functions of these invariants, the remaining conditions of hyperelasticity are included in the model, i.e., normalization and growth conditions, while preserving the polyconvexity of the invariants. Almost twenty years after the work of \textcite{Schroeder2003}, polyconvexity entered the field of data-driven, machine learning-based constitutive modeling, with the first polyconvex physics-augmented neural network (PANN) model introduced by~\textcite{klein2022a}. The basic ideas remained the same: by using polyconvex invariants as arguments for potentials with suitable monotonicity and convexity constraints, polyconvex potentials can be constructed. As proposed by \textcite{Amos2017}, suitable restrictions provide convex and monotonic neural network (NN) architectures, which can then be used as constitutive model equations for polyconvex constitutive models~\parencite{klein2022a}. In recent years, various polyconvex PANN formulations have been proposed using structural tensor-based invariants~\parencite{linden2023,tac2022}, triclinic invariants and group symmetrization~\parencite{klein2022a,klein2026a}, and principal stretches or signed singular values~\parencite{Geuken_Kurzeja_Wiedemann_Mosler_2025,VIJAYAKUMARAN2024106015}.

Polyconvex constitutive models have found application across a wide range of materials, including soft biological tissue~\parencite{LINKA2023134,LINKA2023116007,STPIERRE2023116236,Balzani2006}, rubber-like materials~\parencite{tac2023,klein2025,dammass2025invariantsmatterrolei1,Steinmann_Hossain_Possart_2012}, and homogenized (multiphysical) metamaterials or composites~\parencite{klein2022a,klein2024a,kalina2024a,klein2026a,GEBHART2022111984}. The \textbf{benefits of polyconvex constitutive modeling} are well understood. For a lot of applications, polyconvex constitutive models show an excellent performance in representing the material behavior. More than that, including polyconvexity can significantly improve the model performance in terms of generalization and numerical stability and robustness~\parencite{klein2025,kalina2024a,klein2022a,Schroeder_Neff_Balzani_2005,linden2023}. 


\subsection{What can be done when a polyconvex model performs poorly?}\label{sec:mitigation}

While polyconvexity has proven a useful constitutive condition in many applications, it can be overly restrictive in other scenarios. Then, polyconvex constitutive models show a moderate to poor performance in representing the ground truth material behavior~\parencite{klein2022a,kalina2024a}, practically preventing the use of the considered model. For such scenarios, different \textbf{mitigation strategies} have been proposed.\footnote{Most of the following publications do not explicitly frame their approaches as ``mitigation strategies''. However, as we will demonstrate, they can be interpreted as such.} 

First, when one specific polyconvex formulation fails to accurately represent the considered material behavior, a \textbf{different polyconvex formulation} might perform better. Even when two different approaches are similar in that they satisfy the same constitutive conditions of hyperelasticity, they might differ in their flexibility. This was reported in~\textcite{klein2022a,klein2026a} for models using structural-tensor based invariants or models using triclinic invariants and group symmetrization, in~\textcite{tac2023} for models based on convex NN architectures and models based on neural ordinary differential equations, and in \textcite{wollner2026} for different isotropic model formulations. Moreover, in~\cref{chap:application_micro} of this work, we demonstrate another example for the latter. 

Next, by using \textbf{relaxed convexity conditions}, the constitutive model remains some structure while its flexibility can be increased. For instance, PANN models that are convex in the coordinates of the right Cauchy-Green tensor $\bC$ have been successfully employed for a variety of material behaviors \parencite{asad2022,Zheng_Kochmann_Kumar_2024}. As we discuss in~\textcite[Rem.~4.1]{klein2026a}, convexity of the hyperelastic potential in $\bC$ can be seen as a relaxed polyconvexity condition. In a similar fashion, monotonicity in isotropic invariants was proposed as a relaxed version of polyconvexity in~\textcite{klein2025}. 

Finally, in some scenarios, \textbf{non-polyconvex} constitutive models can be employed, which can further increase model flexibility compared to polyconvex formulations. This is possible when \textbf{suitable calibration data} is available, i.e., datasets featuring a wide range of loading scenarios. This was reported in~\parencite{klein2024a,kalina2025,klein2026a}, and we will show an example for this in~\cref{chap:application_micro} of this work. As proposed by~\textcite{kalina2024a}, the performance of non-polyconvex constitutive models can be promoted by including \textbf{convexity-promoting loss terms} in the model calibration together with a relaxed local version of polyconvexity. This approach is demonstrated in~\textcite{kalina2024a} for homogenization data of magneto-elastic metamaterials and in~\textcite{klein2026a} for experimental data of soft rubber-like materials. 

In principle, constitutive models cannot only learn to predict the ground-truth stress response, but may also adopt the ellipticity of the underlying material behavior from the calibration data -- at least for some deformation modes -- even without explicit polyconvexity constraints. However, non-polyconvex formulations cannot guarantee ellipticity across all deformation scenarios and often exhibit reduced generalization capabilities. Therefore, such formulations should be employed with caution and only when necessary.

\subsection{Main contributions and outline}

While polyconvexity has proven to be a useful constitutive condition in many applications, it can be overly restrictive in other scenarios, leading to limited performance of polyconvex constitutive models and motivating the need for mitigation strategies. This raises the fundamental question:
\begin{center}
\emph{What limits the applicability of polyconvex constitutive models?}
\end{center}

\noindent
Several works have partially addressed this question, covering aspects such as non-polyconvex anisotropic invariants~\parencite{Schroeder2010a,klein2026a}, structural constraints on the energy potential~\parencite{klein2024a}, or the fact that polyconvexity is sufficient but not necessary for ellipticity~\parencite[Sec.~5.3.7]{Dacorogna2008}. However, to the best of our knowledge, a holistic investigation on the limited applicability of polyconvex constitutive models has yet to be conducted. 

The \textbf{main contributions} of this work are as follows.\footnote{Parts of the research presented in this manuscript have been published in the first author's doctoral thesis \parencite{kleinDiss}.} \textbf{(1)} We analyze the theoretical reasons why polyconvexity may, in some cases, impose overly restrictive constraints that limit the achievable accuracy of constitutive models. Thereby, we provide analytical ellipticity guarantees for two non-polyconvex Mooney-Rivlin type potentials. \textbf{(2)} We investigate the practical limitations of polyconvex PANN constitutive models using two representative formulations: models using structural tensor-based invariants and models using signed singular values. Their performance is evaluated on datasets obtained from homogenized microstructured materials, and the accuarcy and reliability of their predictive capabilities are assessed in finite element simulations. \textbf{(3)} Overall, we provide an overview of benefits, limitations, and mitigation strategies of polyconvex hyperelastic constitutive modeling.

The \textbf{outline} of the remaining manuscript is as follows. In~\cref{chap:basics}, we introduce the fundamentals of polyconvex hyperelasticity and PANN constitutive modeling. In~\cref{sec:inv_conv_examples}, we derive analytical ellipticity guarantees for two non-polyconvex Mooney-Rivlin type potentials, and thereby illustrate the structural restrictions of polyconvex invariant-based hyperelasticity. Subsequently, in~\cref{chap:application_micro}, we apply the different PANN models to different datasets of homogenized microstructured materials. Thereby, we investigate both the stress prediction quality of the models and their performance in nonlinear finite element simulations. This is followed by the conclusion in~\cref{sec:conc}. We introduce the notation employed in this work in Appendix~\ref{app:notation}.

\section{Fundamentals of polyconvex hyperelasticity and PANN constitutive modeling}\label{chap:basics}

In this section, the fundamentals of polyconvex hyperelasticity and physics-augmented neural network (PANN) constitutive modeling are introduced. In \cref{sec:kinematics}, the relevant kinematics and constitutive conditions of hyperelasticity are outlined. In~\cref{sec:invs}, we introduce polyconvex constitutive modeling using structural tensor-based invariants. The application thereof for PANN modeling is introduced in in~\cref{sec:PANNs}, follwed by PANN models based on signed singular values in~\cref{sec:PANNsSSV}.

\subsection{Kinematics and constitutive conditions of hyperelasticity}\label{sec:kinematics}

Consider a body in its reference configuration $\mathcal{B}_0\subset\bbR^3$ at the time $t_0\in\bbR$ and its current configuration $\mathcal{B}\subset\bbR^3$ at the time $t\in\mathcal{T}:=\{\tau\in\bbR\,\rvert\,\tau\geq t_0\}$. The mapping $\bphi:\mathcal{B}_0\times\mathcal{T}\rightarrow\mathcal{B}$ links material particles $\bx_0\in\mathcal{B}_0$ to $\bx=\bphi(\bx_0,\,t)\in\mathcal{B}$. Associated with $\bphi$, the \textbf{deformation gradient} $\bF\in\text{GL}^+(3)$, its \textbf{cofactor} $\bH\in\text{GL}^+(3)$, and its \textbf{determinant} $J\in\bbR^+$ are defined as \parencite{Haupt2002,bonet2015}
\begin{equation}\label{eq:def_grad}
\bF=\nabla_0\bphi\,,\qquad \bH=\cof\bF=J\bF^{-T}=\frac{1}{2}\bF\Cross\bF \,,  \qquad  J=\det\bF=\frac{1}{6}\bF:(\bF\Cross\bF)\,.
\end{equation}
The \textbf{isochoric deformation gradient} is defined as $\bar{\bF}=J^{-1/3}\bF$. As an alternative deformation measure, we introduce the \textbf{right Cauchy-Green tensor} $\bC=\bF^T\cdot\bF\in\SYM^+(3)$ and its cofactor $\bG=\cof\bC\in\SYM^+(3)$. The \textbf{principal stretches} $\lambda_i\in\bbR$ are the eigenvalues of the deformation gradient $\bF$, and $\bar{\lambda}_{\beta}$ denote the principal stretches of the isochoric deformation gradient $\bar{\bF}$. The singular values $\sigma_i=\sqrt{\lambda_i^2}\in\bbR_{>0}$ of $\bF$ are the square root of the eigenvalues of the right Cauchy-Green tensor $\bC$. The \textbf{signed singular values} $\nu_i$ have the same absolute value as the singular values, i.e., $\lvert\nu_i\rvert=\sigma_i$, but $\nu_i$ may become negative by rotation, subject to $\nu_1\nu_2\nu_3=\sigma_1\sigma_2\sigma_3$. Note that the principal stretches and signed singular values are isotropic invariants of the right Cauchy-Green tensor $\bC$.

\medskip

In hyperelasticity, we formulate a constitutive model that connects the deformation gradient $\bF$ with the \textbf{first Piola-Kirchhoff stress tensor} $\bP$ at a material point in a body. This is done by introducing a \textbf{hyperelastic potential} $W$ and defining the stress at its gradient field:
\begin{equation}
    W:\GL^+(3)\rightarrow\bbR\,,\quad \bF\mapsto W(\bF)\qquad \text{and}\qquad  \bP=\partial_{\bF}W(\bF)\,.
\end{equation}
By that, \textbf{thermodynamic consistency} is fulfilled by construction. In addition, we consider the following basic constitutive conditions, see e.g. \parencite{Haupt2002,Holzapfel2000}:
\begin{itemize}
\setlength\itemsep{0.2em}
\item \textbf{objectivity}: $W(\bQ\cdot\bF)=W(\bF)\quad\forall(\bF,\,\bQ)\in\GL^+(3)\times\SO(3)$,
\item \textbf{material symmetry}: $W(\bF\cdot\bQ^T)=W(\bF) \quad\forall(\bF,\,\bQ)\in\GL^+(3)\times\mathcal{G}$, \\ where $\mathcal{G}\subseteq\operatorname{O}(3)$ denotes the symmetry group of the considered material,
\item \textbf{stress normalization:} $ \bP(\bF)\big\rvert_{\bF=\bI}=\bnull$,
\item \textbf{volumetric growth condition:} $W(\bF)\rightarrow\infty$ as $J \rightarrow 0^+$.
\end{itemize}

Further constitutive conditions are grounded in the concept of convexity. From an engineering perspective, the most appealing convexity condition is the \textbf{ellipticity} (or \textbf{rank-one convexity}) condition \parencite{Zee1983}
\begin{equation}\label{eq:ellip_mech_comp}
\begin{aligned}
(\ba\otimes\bb):\partial^2 _{\bF\bF}W:(\ba\otimes\bb)\geq 0
\qquad \forall \ba,\bb\in\bbR^3\,,\,\,\bF+\ba\otimes\bb\in\GL^+(3)\,.
    \end{aligned}
\end{equation}
While ellipticity is a very attractive constitutive condition, it is practically impossible to directly formulate hyperelastic potentials that are elliptic by construction \parencite{neff2015}. A sufficient condition for ellipticity is the \textbf{polyconvexity} condition introduced by \textcite{Ball1976,Ball1977}, which can be practically employed in a constitutive modeling framework \parencite{Schroeder2003}. Polyconvex potentials allow for a (non-unique) representation of the hyperelastic potential as
\begin{equation}\label{eq:pc_mech_comp}
    \cP:\cL_2\times\cL_2\times\bbR\rightarrow\bbR\,,\qquad (\bF,\,\bH,\,J)\mapsto\cP(\bF,\,\bH,\,J)\,,
\end{equation}
with the convex function $\cP$ and $W(\bF)=\cP(\bF,\,\bH,\,J)$. We provide an engineering perspective on polyconvexity and ellipticity in~\cref{sec:eng_perspective}.

\medskip

To foster understanding of the relation between aforementioned convexity conditions, we now consider potentials of the form
\begin{equation}\label{eq:pot_ext_comp}
    \tW:\cL_2\times\cL_2\times\bbR\rightarrow\bbR\,,\qquad (\bF,\,\bH,\,J)\mapsto\tW(\bF,\,\bH,\,J)\,,
\end{equation}
with $W(\bF)=\tW(\bF,\,\bH,\,J)$, that explicitly include all arguments of the polyconvexity condition. For sufficiently smooth functions, the Hessian is positive semi-definite (p.s.d.) \parencite{Silhavy2014}, which yields the general convexity condition \parencite{bonet2015}
\renewcommand\arraystretch{1.2}
\begin{equation}\label{eq:general_conv}
\begin{aligned}
\delta\bF:d^2 _{\bF\bF}\tW:\delta\bF
=
\underbrace{\begin{bmatrix}
    \delta\bF: \\
    (\delta\bF\Cross\bF): \\
    \delta\bF:\bH
    \end{bmatrix}
    [\mathbb{H}_{\tW}]
\begin{bmatrix}
    :\delta\bF \\
    :(\delta\bF\Cross\bF) \\
    \delta\bF:\bH
    \end{bmatrix}}_{\text{constitutive type term}}
+\underbrace{\left(\partial_{\bH}\tW+\partial_{J}\tW\,\bF\right):(\delta\bF\Cross\delta\bF) }_{\text{geometric type term}}
    \geq 0 \,,
\end{aligned}
\end{equation}
with the Hessian operator $[\mathbb{H}_{\tW}]$ defined as
\renewcommand\arraystretch{1.4}
\begin{equation}\label{eq:hessian_mech}
   [\mathbb{H}_{\tW}]:=  \begin{bmatrix}
   \partial^2_{\bF\bF} \tW&\partial^2_{\bF\bH}\tW & \partial^2_{\bF J}\tW\\
\partial^2_{\bH\bF}\tW & \partial^2_{\bH\bH}\tW & \partial^2_{\bH J}\tW\\
\partial^2_{J\bF }\tW& \partial^2_{J \bH }\tW & \partial^2_{JJ}\tW
\end{bmatrix}  \,.
\end{equation}
The first term in~\cref{eq:general_conv} includes second derivatives of the potential $\tW$ w.r.t. $(\bF,\,\bH,\,J)$, which suggests to phrase it as a ``constitutive type term'', while the second term includes first derivatives of the potential w.r.t. $(\bH,\,J)$, which suggests to phrase it as a ``geometric type term'' \parencite{poya2024}. For polyconvex potentials, the Hessian operator $[\mathbb{H}_{\tW}]=[\mathbb{H}_{\cP}]$ is p.s.d., and the constitutive type term in \cref{eq:general_conv} takes only positive values. However, the geometric type term can still take negative values, from which it becomes apparent that polyconvexity does not imply convexity of the potential in $\bF$ alone. For ellipticity, we have $\delta\bF=\ba\otimes\bb$, and the geometric type term in \cref{eq:general_conv} vanishes due to $(\ba\otimes\bb)\Cross(\ba\otimes\bb)=\boldsymbol{0}$ \parencite{bonet2015}. Thus, polyconvexity implies ellipticity. Note that polyconvexity is sufficient but not necessary for ellipticity, which we further elaborate on in~\cref{sec:inv_conv_examples}.

\subsection{Polyconvex constitutive modeling using structural tensor-based invariants}\label{sec:invs}

By formulating a hyperelastic potential in terms of invariants of $\bC$ and tuples of structural tensors, material symmetry and objectivity can be fulfilled by construction \parencite{Holzapfel2000,riemer2025}. The isotropicization theorem states that by including structural tensors in the arguments of the hyperelastic potential, it can be expressed as an isotropic tensor function even if the considered material behavior is anisotropic \parencite{itskov2015,Haupt2002}. Representation theorems for scalar-valued tensor functions provide expressions for invariants of the right Cauchy-Green tensor $\bC$ and the tuple of structural tensors $\cS^{\square}$ \parencite{boehler1977,Xiao1996OnIE}, where $\square$ denotes the considered symmetry group. Based on this, different tuples of invariants can be derived. Let $\bcI^{\square}:=(J_1^{\square},\,\dotsc,\,J_m^{\square})\in\bbR^m$ be an $m$-tuple containing invariants. We call the tuple of invariants an \emph{integrity basis} when any invariant of the considered symmetry group can be represented as a polynomial in the invariants contained in $\bcI^{\square}$. Further, we call the tuple of invariants a \emph{functional basis} when any invariant of the considered symmetry group can be represented as a function of the invariants contained in $\bcI^{\square}$. Every integrity basis is a functional basis but not vice versa. For a comprehensive introduction to these concepts, the reader is referred to~\textcite{riemer2025}. 

\medskip

In invariant-based modeling, the hyperelastic potential does not directly depend on the deformation gradient $\bF$. Instead, the potential $W(\bF)=\psi(\bcI^{\square})$ is a function of a set of invariants, i.e.,
\begin{equation}\label{eq:pot_inv_basics}
    \psi:\bbR^m\rightarrow\bbR\,,\quad \bcI^{\square}\mapsto \psi(\bcI^{\square})\qquad \text{and}\qquad  \bP=\partial_{\bF}W(\bF)=\sum_{\alpha=1}^m\partial_{\cI^{\square}_{\alpha}}\psi\,d_{\bF}\cI^{\square}_{\alpha}\,.
\end{equation}
The derivatives of the potential w.r.t.~the invariants are referred to as \emph{stress coefficients}, while the derivatives of the invariants w.r.t.~the deformation gradient are referred to as \emph{tensor generators} \parencite{kalina2022}. Polyconvexity of the potential in~\cref{eq:pot_inv_basics} is equivalent to the p.s.d. of the Hessian
\begin{equation}\label{eq:hess_inv}
d^2_{\bxi\bxi}\psi(\bcI^{\square})=[\mathbb{H}_{\psi}]=\underbrace{\left(\partial_{\bxi}\bcI^{\square}\right)^T\cdot\partial^2_{\bcI^{\square}\bcI^{\square}}\psi\cdot\left(\partial_{\bxi}\bcI^{\square}\right)}_{\text{constitutive type term}}\,\,+\underbrace{\partial_{\bcI^{\square}}\psi\cdot\partial^2_{\bxi\bxi}\bcI^{\square}}_{\text{geometric type term}}\succeq 0\,,
\end{equation}
where $\bxi:=(\bF,\,\bH,\,J)\in\bbR^{19}$ are the arguments of the polyconvexity condition (cf.~\cref{eq:pc_mech_comp}). Since, in general, the invariants contained in $\bcI^{\square}$ are nonlinear functions of $(\bF,\,\bH,\,J)$, the Hessian operator $[\mathbb{H}_{\psi}]$ consists of a constitutive and a geometric type term \parencite{poya2024}. In constitutive modeling practice, it is infeasible to consider p.s.d.~of both terms individually. Rather, p.s.d.~of both the constitutive type term and the geometric type term individually is applied \parencite{klein2022a,Schroeder2003}, which is a sufficient but not necessary condition for $[\mathbb{H}_{\psi}]$ to be p.s.d. This results in the sufficient conditions that (i) the potential $\psi$ is a convex and monotonic\footnote{In this work, if not stated otherwise, ``monotonic'' refers to component-wise monotonically increasing functions, i.e., $\partial_{x_i}f(\bx)\geq 0$.} function in the invariants $\bcI^{\square}$ and that (ii) all invariants contained in $\bcI^{\square}$ are polyconvex.
\begin{center}
\fbox{\parbox{0.98\linewidth}{
The conditions for polyconvexity of potentials using structural tensor-based invariants are \textbf{sufficient but not necessary}, resulting in overly restrictive constitutive models that cannot represent every polyconvex function.}}
\end{center}

\noindent
 When constructing invariant sets based on representation theorems, the polyconvexity condition is not considered, and the obtained invariants are generally not polyconvex. Thus, the integrity bases obtained from representation theorems have to be adapted to obtain polyconvex integrity or functional bases applicable for polyconvex constitutive modeling, which we describe in detail in \textcite[Sec.~3]{klein2026a}. For instance, for the isotropic symmetry group $\cK$, the structural tensor is the second-order identity tensor ($\cS^{\text{iso}}=(\bI)$), and a polyconvex integrity basis can be derived as $\bcI^{\text{iso}}=\big(\IoneISO,\,\ItwoISO,\,\IthreeISO\big)$ with
\begin{equation}\label{eq:invs_iso}
\begin{aligned}
    \IoneISO = \tr\bC=\norm{\bF}^2\,,\qquad \ItwoISO=\tr\bG=\norm{\bH}^2\,,\qquad \IthreeISO=\det\bC=J^2\,.
    \end{aligned}
\end{equation}
For the transversely isotropic symmetry group $\mathcal{D}_{\infty}$, the single second-order structural tensor $\bMti =\bn_0\otimes\bn_0$ is sufficient to characterize the materials anisotropy ($\cS^{\text{ti}}=(\bMti)$), where $\bn_0$ denotes the unit preferred direction in the reference configuration. A polyconvex integrity basis for transverse isotropy can be derived as  $ \bcI^{\text{ti}}=\big(\IoneISO,\,\ItwoISO,\,\IthreeISO,\,\IoneTI,\,\ItwoTI\big)$~\parencite{Schroeder2003}, with the isotropic invariants from~\cref{eq:invs_iso} and
\begin{equation}\label{eq:invs_ti_pc}
\IoneTI=\tr\bC\cdot\bMti=\norm{\bF\cdot\bMti}^2\,,\qquad \ItwoTI=\tr\bG\cdot\bMti=\norm{\bG\cdot\bMti}^2\,.
\end{equation}
In~\textcite{klein2026a}, we provide polyconvex integrity and functional bases for five crystal groups, one transversely isotropic group, and the isotropic group. However, six more crystal groups and one more transversely isotropic group are also relevant in hyperelastic constitutive modeling~\parencite{riemer2025}. To the best of our knowledge, no polyconvex integrity or functional bases have been proposed for these remaining symmetry groups. The formulation of polyconvex bases for all symmetry groups entails challenges such as invariants that include higher-order powers of $\bC$, structural tensors that are not p.s.d., and complex fourth- and sixth-order structural tensors. 
\begin{center}
\fbox{\parbox{0.98\linewidth}{
For a lot of symmetry groups, \textbf{no polyconvex integrity or functional bases} are available. Thus, polyconvex constitutive models for such symmetry groups must rely on incomplete sets of invariants. This leads to a loss of information and restricts the expressive capability of the resulting models. More precisely, it is generally not possible to represent arbitrary invariant functions, in particular the hyperelastic potential, solely in terms of the elements of such sets.
}}
\end{center}

\subsection{Polyconvex PANN constitutive models using structural tensor-based invariants}\label{sec:PANNs}

In the literature, different PANN constitutive model formulations have been proposed, where the most commonly employed approaches use \textbf{structural tensor-based invariants}. Polyconvex invariant-based PANN models were first proposed by \textcite{klein2022a} and later extended by invariant-based stress normalization terms by \textcite{linden2023}, resulting in a model that fulfills all common constitutive conditions of hyperelasticity by construction. The overall PANN model is expressed as
\begin{equation}\label{eq:PANN}
    \psi^{\text{PANN}}_{\square}(\bcI^{\square})=\psi^{\text{NN}}_{\square}(\bcI^{\square})+\psi^{\text{stress}}_{\square}(\bcI^{\square})+\psi^{\text{growth}}(J)\,.
\end{equation}
Here, $\psi^{\text{NN}}_{\square}$ is the neural network part of the potential, for which feed-forward neural networks (FFNNs) are employed. In this work, we consider FFNNs of the form
\begin{equation}\label{eq:FFNN_1}
\begin{aligned}
{\mathrlap{\bx^{(h)}}\phantom{\bx^{(h)}}}&=\SP\big(\bW^{(h)}\cdot\bx^{(h-1)}+\bb^{(h)}\big)&&\in\bbR^{n_h}\,,\quad h=1,\,\dotsc H \,,
\\
{\mathrlap{\psi^{\text{NN}}_{\square}}\phantom{\psi^{\text{NN}}_{\square}}}&=\bw^{(H+1)}\cdot\bx^{(H)}&&\in\bbR\,,
\end{aligned}
\end{equation}
with $\bx^{(0)}=\bcI^{\square}$ and $n_0=m$. The layers from 1 to $H$ are commonly referred to as hidden layers, while the layer $H+1$ is referred to as output layer. The number of layers $H+1$ denotes the depth of the NN, which is considered as a hyperparameter, together with the number of nodes in each layer $n_h$. The weight matrices $\bW^{(h)}\in\bbR^{n_{h}\times n_{h-1}},\bw^{(H+1)}\in\bbR^{n_{H}}$ and the bias vectors $\bb^{(h)}\in\bbR^{n_h}$ form the set of parameters $\boldsymbol{\mathcal{P}}$ that are optimized to fit the NN to a given dataset. The component-wise applied softplus activation function is denoted as $\SP(x)=\log(1+e^x)$, which we employ as it is a convex and monotonic function. Note that we employ a linear activation function and no bias in the output layer. As originally proposed by \textcite{Amos2017}, the simple structure and recursive definition of \cref{eq:FFNN_1} makes FFNNs a very natural choice for the construction of monotonic and convex functions. When all weights are non-negative, the NN in \cref{eq:FFNN_1} is convex and monotonic, which refer to as convex-monotonic NNs (CMNNs) \parencite{Amos2017}. Thus, when all invariants in $\bcI^{\square}$ are polyconvex, the monotonic and convex NN potential $ \psi^{\text{NN}}_{\square}$ is polyconvex. The NN potential is complemented by the normalization term $\psi^{\text{stress}}_{\square}$ to ensure stress normalization and the volumetric growth term $\psi^{\text{growth}}$ to ensure fulfillment of the volumetric growth condition. A detailed introduction of these terms and the required polyconvex invariants for $\bcI^{\square}$ can be found in \textcite{linden2023} and \textcite{klein2026a}.

\subsection{Polyconvex PANN constitutive models based on signed singular values}\label{sec:PANNsSSV}

For isotropic hyperelasticity, polyconvex constitutive models \textbf{based on signed singular values} have recently been proposed as an alternative to formulations based on the main invariants of $\bC$ \parencite{Geuken_Kurzeja_Wiedemann_Mosler_2025}. This approach is based on recent theoretical results by~\textcite{wiedemann2023characterizationpolyconvexisotropicfunctions}. 

For this framework, we consider the hyperelastic potential $W(\bF)=\Psi(\nu_1,\,\nu_2,\,\nu_3)$ with 
\begin{equation}\label{eq:pot_ss}
  \Psi:\bbR^{3}\rightarrow\bbR\,, \qquad   (\nu_1,\,\nu_2,\,\nu_3)\mapsto \Psi (\nu_1,\,\nu_2,\,\nu_3)\,,
\end{equation}
based on the signed singular values $\nu_i$ (cf.~\cref{sec:kinematics}). Objectivity and isotropy imply that $\Psi$ is $\Pi_3$-invariant with
\begin{equation}
    \Pi_3=\left\{\boldsymbol{B}\cdot\operatorname{diag}(\epsilon)\,\rvert\,\boldsymbol{B}\in\text{Perm}(3),\,\epsilon\in\{-1,1\}^3,\,\epsilon_1\epsilon_2\epsilon_3=1\right\}\,.
\end{equation}
This includes four symmetries and six permutations
\begin{equation}
\begin{aligned}
    \Psi (\nu_1,\,\nu_2,\,\nu_3)=\Psi (-\nu_1,\,-\nu_2,\,\nu_3)=\Psi (-\nu_1,\,\nu_2,\,-\nu_3)=\Psi (\nu_1,\,-\nu_2,\,-\nu_3)\,,
    \\
        \Psi (\nu_1,\,\nu_2,\,\nu_3)=\Psi (\nu_1,\,\nu_3,\,\nu_2)=\Psi (\nu_2,\,\nu_1,\,\nu_3)=\Psi (\nu_2,\,\nu_3,\,\nu_1)=\Psi (\nu_3,\,\nu_1,\,\nu_2)=\Psi (\nu_3,\,\nu_2,\,\nu_1)\,,
\end{aligned}
\end{equation}
which, combined, provide the 24 elements of $\Pi_3$ \parencite[Table~1]{Geuken_Kurzeja_Wiedemann_Mosler_2025}. The $\Pi_3$-invariant potential $\Psi$ is polyconvex if and only if there exists a representation $ \Psi (\nu_1,\,\nu_2,\,\nu_3)= \widetilde{\Psi} (\boldsymbol{\nu})$, where 
\begin{equation}\label{eq:pot_princ_stretch}
\begin{aligned}
    \widetilde{\Psi}:\bbR^{7}&\rightarrow\bbR\,,  \qquad\boldsymbol{\nu}\mapsto \widetilde{\Psi} (\boldsymbol{\nu})\quad\text{with}\quad\boldsymbol{\nu}=(\nu_1,\,\nu_2,\,\nu_3,\,\nu_1\nu_2,\,\nu_1\nu_3,\,\nu_2\nu_3,\,\nu_1\nu_2\nu_3)\,,
\end{aligned}
\end{equation}
is a convex function \parencite{wiedemann2023characterizationpolyconvexisotropicfunctions}. 

This formulation can be transferred to the PANN model~\parencite{Geuken_Kurzeja_Wiedemann_Mosler_2025}
\begin{equation}\label{eq:PANN_SSV}
    \Psi^{\text{PANN}}(\boldsymbol{\nu})=\frac{1}{24}\sum_{\bB\in\Pi_3}\Big[
     \Psi^{\text{NN}}(\bB\cdot\boldsymbol{\nu})\Big]+\Psi^{\text{stress}}(J)
    +\Psi^{\text{growth}}(J)\,.
\end{equation}
When $\Psi^{\text{NN}}$ is represented by an input-convex neural network, i.e., if the output of the FFNN in~\cref{eq:FFNN_1} is convex in its input, the NN potential is polyconvex. For~\cref{eq:FFNN_1}, this can be fulfilled when all weights, except those in the first hidden layer, are non-negative. In~\cref{eq:PANN_SSV}, the summation over the elements of the $\Pi_3$ group is performed to ensure that the potential $\Psi^{\text{PANN}}$ is invariant w.r.t.~the action of this group. A detailed introduction of the stress normalization term $\Psi^{\text{stress}}$ and the volumetric growth term $\Psi^{\text{growth}}$ can be found in~\textcite{Geuken_Kurzeja_Wiedemann_Mosler_2025}.

\begin{remark}
While the principal stretches $\lambda_i$ and the signed singular values $\nu_i$ appear similar at first sight (cf.~\cref{sec:kinematics}), the theoretical results on polyconvexity obtained for them differ significantly. Polyconvex constitutive models based on principal stretches have already been established in Ball's original work on polyconvexity \parencite{Ball1976}. However, his formulation provides \emph{sufficient but not necessary} conditions and is therefore not guaranteed to represent every polyconvex isotropic function. This limitation is similar to the one encountered in models using structural tensor-based invariants. In contrast, \textcite{wiedemann2023characterizationpolyconvexisotropicfunctions} recently derived \emph{sufficient and necessary} conditions for polyconvex isotropic hyperelasticity based on signed singular values applicable to practical constitutive modeling, see also \parencite{Rosakis1997, Mielke2005}. Consequently, with a suitable NN architecture, PANN models based on signed singular values have universal approximation properties for isotropic polyconvex hyperelasticity~\parencite{Geuken_Kurzeja_Wiedemann_Mosler_2025}. Thus, signed singular value-based models offer more flexibility than principal stretch-based models, which is the reason why we employ the former.
\end{remark}

\section[Analytical examples illustrating the structural restrictions of polyconvex invariant-based hyperelasticity]{Analytical examples illustrating the structural restrictions \\ of polyconvex invariant-based hyperelasticity}\label{sec:inv_conv_examples}

Building on the theory of polyconvex hyperelasticity using structural tensor-based invariants introduced in the previous section, we now further examine the structural restrictions inherent to this modeling framework. In~\cref{sec:invs}, we have already discussed that for some anisotropy classes, the function space that polyconvex constitutive models using structural tensor-based invariants can represent is restricted since no polyconvex integrity or functional basis is available. However, even if a polyconvex integrity or functional basis is available for the considered symmetry group, the constitutive model is subject to further structural constraints. In particular, in~\cref{sec:invs}, we derived monotonicity and convexity of the potential as sufficient but not necessary conditions for polyconvexity.

\begin{figure}[t!]
\centering
\includegraphics[width=\textwidth]{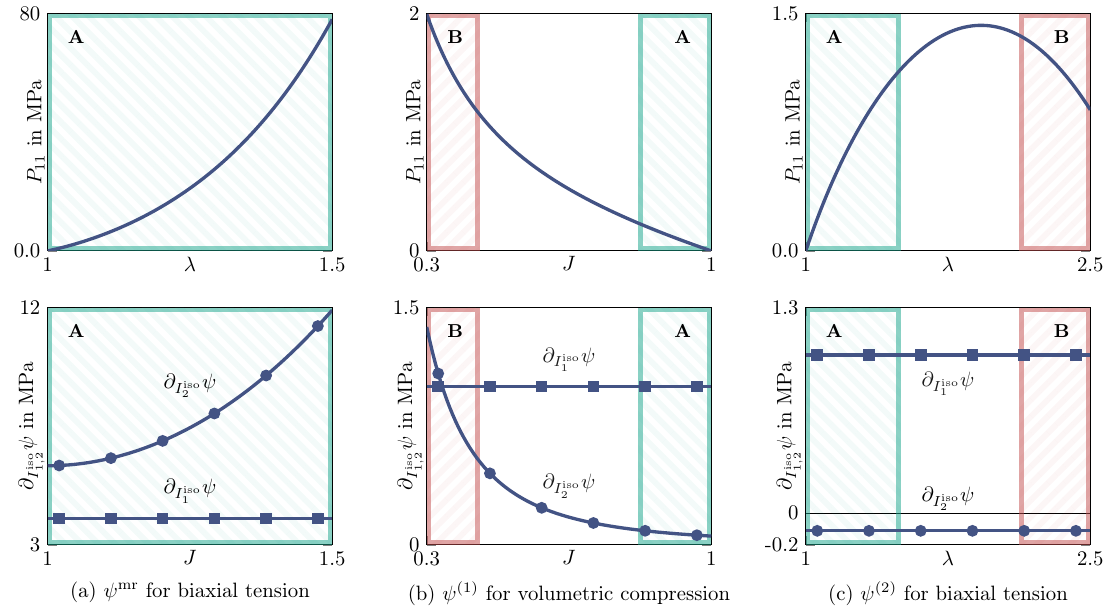}
\caption{Investigation of analytical potentials. Area A indicates an analytical ellipticity guarantee, while area B indicates a numerically evaluated loss of ellipticity. \textbf{(a)} $\psi^{\text{mr}}$ is convex and monotonic in the invariants, and thus polyconvex (and elliptic) by construction. \textbf{(b)} $\psi^{(1)}$ is concave in $\ItwoISO$, which can lead to a loss of ellipticity. \textbf{(c)} $\psi^{(2)}$ is monotonically decreasing in $\ItwoISO$, which can lead to a loss of ellipticity.} 
\label{fig:ana_examples}
\end{figure}

As an example for such a polyconvex potential, let us consider the isotropic Mooney-Rivlin type potential\footnote{Note that we do not consider the energy normalization condition $W(\bI)=0$ in this work, cf.~\textcite[Rem.~2.1]{klein2026a}.}
\begin{equation}
\psi^{\text{mr}}=4\mu_1\IoneISO+\mu_2(\ItwoISO)^2-2(4\mu_1+12\mu_2)J\,.
\end{equation} 
with $\mu_1=\mu_2=1\text{ MPa}$. This potential is convex and monotonically increasing in $(\IoneISO,\,\ItwoISO)$ and convex in $J$, where the associated Hessian operator given by
\renewcommand\arraystretch{1.2}
\begin{equation}\label{eq:hessian_mech_example_pc}
   [\mathbb{H}_{\psi^{\text{mr}}}]= 8 \begin{bmatrix}
   \bnull&\bnull& \bnull \\
\bnull& \mu_2\,\bH\otimes\bH & \bnull \\
\bnull& \bnull & \bnull
\end{bmatrix}+
4 \begin{bmatrix}
   2\mu_1 \bbI&\bnull& \bnull \\
\bnull& \mu_2\ItwoISO\bbI & \bnull \\
\bnull& \bnull & \bnull
\end{bmatrix}\,,
\end{equation}
as a special case of the general Hessian operator for invariant-based potentials in~\cref{eq:hess_inv}. Since $\ItwoISO>0$ \parencite{Schroeder2003}, both terms in~\cref{eq:hessian_mech_example_pc} are p.s.d., and consequently, the overall Hessian $[\mathbb{H}_{\psi^{\text{mr}}}]$ is p.s.d. In~\cref{fig:ana_examples}(a), the behavior of $\psi^{\text{mr}}$ for a biaxial deformation state $\bF=\operatorname{diag}(\lambda,\,\lambda,\,\lambda^{-2})$ is investigated, where the stress and the stress coefficients are both visualized. Both stress coefficients are positive and non-decreasing along the considered deformation path, which aligns with the Hessian operator in~\cref{eq:hessian_mech_example_pc}.

\medskip

As discussed in~\cref{sec:invs}, p.s.d. of both individual terms in the Hessian is sufficient but not necessary for polyconvexity. Next, recall that, from an engineering perspective, polyconvexity is desirable not for its own sake, but rather to ensure ellipticity of the constitutive model (cf.~\cref{sec:eng_perspective}).
To recapitulate the relation between both conditions, we express them as (cf.~\cref{eq:general_conv})
\renewcommand\arraystretch{1.2}
\begin{equation}\label{eq:ellip_mech_comp_extended_pc}
\begin{aligned}
\underbrace{
\begin{bmatrix}
    \delta\bF: \\
    \delta\bH: \\
    \delta J
    \end{bmatrix}
    [\mathbb{H}_{\psi}]
\begin{bmatrix}
    : \delta\bF \\
    : \delta\bH \\
    \delta J
    \end{bmatrix}
    \geq 0 }_{\text{polyconvexity}}\,,\qquad\underbrace{\begin{bmatrix}
    (\ba\otimes\bb): \\
    ((\ba\otimes\bb)\Cross\bF): \\
    (\ba\otimes\bb):\bH
    \end{bmatrix}
    [\mathbb{H}_{\psi}]
\begin{bmatrix}
    :(\ba\otimes\bb) \\
    :((\ba\otimes\bb)\Cross\bF) \\
    (\ba\otimes\bb):\bH
    \end{bmatrix}
    \geq 0 }_{\text{ellipticity}}\,.
    \end{aligned}
\end{equation}
For polyconvexity, the inequality must hold for generic values of $\delta\bF$, $\delta\bH$, and $\delta J$, which means that $[\mathbb{H}_{\psi}]$ is p.s.d. In contrast, for ellipticity, the contraction of the Hessian $[\mathbb{H}_{\psi}]$ solely depends on a rank-one tensor $\ba\otimes\bb$ with $\ba,\bb\in\bbR^3$, and the current deformation state through $\bF$ and $\bH$. Thus, p.s.d. of $[\mathbb{H}_{\psi}]$ (and thus polyconvexity) is only sufficient but not necessary for ellipticity. 

To further illustrate these structural restrictions, we now provide analytical ellipticity guarantees within some deformation range for two non-polyconvex potentials. In particular, in the following examples we employ potentials that do not satisfy the monotonicity and convexity constraints of the potential introduced as sufficient conditions for polyconvexity. For the following calculations, we make use of some basic properties of the tensor cross product \parencite{bonet2015,deBoer1982}
\begin{equation}\label{eq:cross_prod_1}
\begin{aligned}
\norm{\bA\Cross\bB}^2&=\phantom{2}\norm{\bA}^2\norm{\bB}^2+\left(\bA:\bB\right)^2-\norm{\bA^T\cdot\bB}^2-\norm{\bA\cdot\bB^T}^2\,,
\\
 \norm{\bA\Cross\bF}^2&\leq 2 \norm{\bA}^2\norm{\bF}^2-2\lambda_{\text{min}}(\bC)\norm{\bA}^2\,.
\end{aligned}
\end{equation}
Note that \cref{eq:cross_prod_1}$_2$ follows from combining \cref{eq:cross_prod_1}$_1$ with the Cauchy-Schwarz inequality $\left(\bA:\bB\right)^2\leq \norm{\bA}^2\norm{\bB}^2$ and the inequality $\norm{\bF\cdot\bA}^2\geq\lambda_{\text{min}}(\bC)\norm{\bA}^2$ \parencite[Lemma~A.3]{Schroeder2003}, where $\lambda_{\text{min}}(\bC)$ denotes the smallest eigenvalue of $\bC$.

\subsection{Compensation of negative eigenvalues between the terms of the Hessian}\label{sec:hessian}

Let us now consider the potential
\begin{equation}\label{eq:ana_pot_ex_1}
    \psi^{(1)}=\mu_1\IoneISO-\frac{\mu_2}{2} (\ItwoISO)^{-1}-2\left(\mu_1+ \frac{\mu_2}{9}\right)J\,,
    \end{equation}
with $ \mu_1=\mu_2=1\text{ MPa}$ and the associated Hessian operator
\renewcommand\arraystretch{1.2}
\begin{equation}\label{eq:hessian_inv_ex_1}
[{\mathbb{H}}_{\psi^{(1)}}]=4
\begin{bmatrix}
\bnull  & \bnull & \bnull \\
\bnull & -\mu_2 (\ItwoISO)^{-3}\,\bH\otimes \bH   & \bnull \\
\bnull & \bnull & \bnull
 \end{bmatrix}
 +
\begin{bmatrix}
2\mu_1\mathbb{I} & \bnull & \bnull\\
\bnull & \mu_2 (\ItwoISO)^{-2}\,\mathbb{I}  & \bnull \\
\bnull & \bnull & \bnull
 \end{bmatrix}\,.
\end{equation}
Note that the constitutive type term of this Hessian operator is negative semi-definite. However, as elaborated above, ellipticity of $\psi^{(1)}$ does not require the p.s.d.\ of $[{\mathbb{H}}_{\psi^{(1)}}]$, let alone p.s.d.\ of its individual contributions. Rather, the ellipticity condition in~\cref{eq:ellip_mech_comp_extended_pc} must hold, which provides the following inequality as a condition for $\psi^{(1)}$ to be elliptic:
\begin{equation}\label{eq:psi2_ellip}
    \begin{aligned}
      &  \norm{\rone}^2+ 0.5 (\ItwoISO)^{-2}\norm{\roneB\Cross\bF}^2-2 (\ItwoISO)^{-3}\big(\bH:(\roneB\Cross\bF)\big)^2
        \\  
         \geq   &\norm{\rone}^2+ 0.5 (\ItwoISO)^{-2}\norm{\roneB\Cross\bF}^2-2 (\ItwoISO)^{-3}\norm{\bH}^2\norm{\roneB\Cross\bF}^2
        \\  
         \geq &\norm{\rone}^2\Big(1- 3 (\ItwoISO)^{-2}\big(\IoneISO-\lambda_{\text{min}}(\bC)\big)\Big)\,.
    \end{aligned}
\end{equation}
From the last row of~\cref{eq:psi2_ellip}, we obtain an ellipticity guarantee for the potential $\psi^{(1)}$ within
\begin{equation}\label{eq:psi_2_ellip_estimate}
\frac{3\big(\IoneISO-\lambda_{\text{min}}(\bC)\big)}{(\ItwoISO)^{2}}\leq 1\,,
\end{equation}
which does not depend on the test vectors $\ba$ and $\bb$. Even if one of the individual terms of the Hessian operator has negative eigenvalues, the overall material behavior can still be elliptic within a certain range of deformations. In our example, the constitutive type term has negative eigenvalues, which are compensated by positive eigenvalues of the geometric type term. It is straightforward to construct a reversed example. In \textcite[Appendix~B]{VIJAYAKUMARAN2024106015}, a similar example is provided. There, a polyconvex potential in terms of the principal stretches is considered. It is shown that the same potential expressed in terms of invariants does not fulfill the sufficient conditions for polyconvexity, in particular, it is not convex in the invariants.

\medskip

We now demonstrate that $\psi^{(1)}$ is not globally elliptic, and thus cannot be polyconvex. For this, we consider the volumetric deformation gradient $\bF=J^{1/3}\bI$ and the test vectors $\ba=\bb=\be^{(1)}$, where $\be^{(1)}$ is the first basis vector of the Cartesian coordinate system. Inserting this into the ellipticity condition in~\cref{eq:psi2_ellip}$_1$ yields
\begin{equation}\label{eq:ellip_ex_1_limit}
 \normm{\be^{(1)}\otimes\be^{(1)}}^2+ \frac{1}{18J^{8/3}}
 \normm{J^{1/3}(\be^{(1)}\otimes\be^{(1)})\Cross\bI}^2- \frac{2}{27J^{4}}\big(J\,\bI:((\be^{(1)}\otimes\be^{(1)})\Cross\bI)\big)^2
=
1-\frac{5}{27J^2}\,,
\end{equation}
which becomes negative as $J\to0^+$. Thus, the potential $\psi^{(1)}$ loses ellipticity for large volumetric compressions.
To further illustrate this, the stress and the stress coefficients of the potential $\psi^{(1)}$ are both visualized in~\cref{fig:ana_examples}(b). For volumetric compression, the analytical estimate in~\cref{eq:psi_2_ellip_estimate} guarantees ellipticity of $\psi^{(1)}$ for $J\geq\sqrt{2/3}\approx 0.82$. A numerical analysis\footnote{The ellipticity condition~\cref{eq:ellip_mech_comp} can be numerically evaluated by checking p.s.d.\ of the acoustic tensor $Q_{ij}=\bbA_{i\alpha j\beta}a_{\alpha}a_{\beta}$, cf.\ \parencite[Sec.~5]{Schroeder_Neff_Balzani_2005}, where $\bbA=\partial^2_{\bF\bF}W$. For this, discrete samples of $\ba\in\bbR^3$ are employed.} reveals a loss of ellipticity for $J\approx 0.43$, which aligns with the critical value of $J=\sqrt{5/27}$ received through the exemplary analytical analysis in~\cref{eq:ellip_ex_1_limit}. Thus, the estimate for the area within $\psi^{(1)}$ is elliptic in~\cref{eq:psi_2_ellip_estimate} is relatively conservative for this deformation mode.
The stress coefficient for $\partial_{\ItwoTI}\psi^{(1)}$ is positive and decreasing, which indicates a positive first derivative and a negative second derivative of the potential $\psi^{(1)}$ w.r.t.\ $\ItwoTI$ (cf.~\cref{eq:hessian_inv_ex_1}). This negative second derivative can cause a loss of ellipticity. While for small volumetric compression, the negative slope of $\partial_{\ItwoTI}\psi^{(1)}$ is moderate, it gets more pronounced as $J$ decreases. Consequently, at some point, it cannot be compensated anymore by the p.s.d.\ contribution of the Hessian in~\cref{eq:hessian_inv_ex_1}, and the potential $\psi^{(1)}$ loses ellipticity. This demonstrates that, see also \parencite{Ghiba_Neff_Martin_2015,kalina2024a}:
\begin{center}
\fbox{\parbox{0.98\linewidth}{
Ellipticity might only be fulfilled \textbf{locally}, whereas polyconvexity is a global condition supposed to hold even for unreasonably large deformations not encountered in engineering practice.}}
\end{center}

\subsection{A Hessian with negative eigenvalues}

Next, we consider the potential
    \begin{equation}
        \psi^{(2)}=\mu_1\IoneISO-\frac{\mu_2}{9} \ItwoISO-2\left(\mu_1-\frac{2\mu_2}{9}\right) J\,,
    \end{equation}
    with $\mu_1=\mu_2=1\text{ MPa}$ and the associated Hessian operator
        \renewcommand\arraystretch{1.2}
\begin{equation}\label{eq:hessian_inv_ex}
[{\mathbb{H}}_{\psi^{(2)}}]=
 2
\begin{bmatrix}
\mu_1\mathbb{I} & \bnull & \bnull\\
\bnull & -\frac{\mu_2}{9}  \,\bbI & \bnull \\
\bnull & \bnull & \bnull
 \end{bmatrix}\,.
\end{equation}
Notably, this Hessian operator is indefinite, i.e., it has positive and negative eigenvalues. From \cref{eq:ellip_mech_comp_extended_pc}, it follows that ellipticity of the potential $\psi^{(2)}$ is equivalent to
\begin{equation}
    \begin{aligned}
        \norm{\rone}^2- \frac{1}{9} \norm{\roneB\Cross\bF}^2
           \geq  \norm{\rone}^2\left(1-\frac{2}{9}\left[I_1-\lambda_{\text{min}}(\bC)\right]\right)\,.
    \end{aligned}
\end{equation}
Thus, we have an ellipticity guarantee for the potential $\psi^{(2)}$ within
\begin{equation}\label{eq:ana_estimate_ex_2}
\IoneISO-\lambda_{\text{min}}(\bC)\leq 4.5\,,
\end{equation}
which demonstrates that the Hessian matrix must not be p.s.d.\ for ellipticity. It is straightforward to demonstrate that $\psi^{(2)}$ is not globally elliptic (and thus not polyconvex), e.g., for volumetric dilatation.
We further investigate the behavior of the potential $\psi^{(2)}$ for a biaxial deformation state $\bF=\operatorname{diag}(\lambda,\,\lambda,\,\lambda^{-2})$, where the stress and the stress coefficients are both visualized in~\cref{fig:ana_examples}(c). For this deformation state, the analytical estimate in~\cref{eq:ana_estimate_ex_2} guarantees ellipticity of $\psi^{(2)}$ for $\lambda\leq 1.5$. A numerical analysis reveals a loss of ellipticity for $\lambda\approx 2.1$. For moderate tension, the $P_{11}$ stress response is increasing. At $\lambda\approx 1.9$, the stress response starts to decrease as the deformation increases, and shortly afterwards, the potential $\psi^{(2)}$ loses ellipticity. This again demonstrates that ellipticity might only be fulfilled locally.

\medskip

These examples illustrate that ellipticity does not require the Hessian operator $[{\mathbb{H}}_{\psi}]$ or its individual contributions to be p.s.d:
\begin{center}
\fbox{\parbox{0.98\linewidth}{
Polyconvexity (i.e., global p.s.d. of the Hessian operator $[\bbH_{\psi}]$) is \textbf{sufficient, but not necessary}\\ for ellipticity.
}}
\end{center}
Moreover, ellipticity might only be fulfilled locally. Neither of this is acknowledged in invariant-based polyconvex constitutive models, where both the constitutive type term and the geometric type term of the Hessian are restricted to be p.s.d.\ for all deformation modes.
As a result of these overly restrictive conditions, polyconvex constitutive models might fail to accurately represent the behavior of a material even within its elliptic regime. 

\begin{remark}
In general, the method applied above can also provide elliptic invariants. For instance, one could employ
\begin{equation}
\tilde{I}^{\text{iso}}_1=\IoneISO-\frac{1}{9} \ItwoISO\,,
\end{equation}
to construct a constitutive model that is elliptic within above given ellipticity guarantee. However, we could not obtain elliptic invariants that improved the prediction quality of our PANN models. For instance, when investigating transversely isotropic invariants, we could not ensure reasonably large deformation areas for the ellipticity guarantee. Still, our investigations provide an interesting illustration of the relation between polyconvexity and ellipticity.
\end{remark}

\section{Application to homogenization data of microstructured materials}\label{chap:application_micro}

In this section, we apply different PANN constitutive models to synthetic homogenization data of microstructured materials. In~\cref{sec:data_models}, we provide information on the considered datasets, PANN models, and the calibration strategy. We conduct a hyperparameter study in~\cref{sec:application_micro_calib}. This is followed by an evaluation of the calibrated PANN models in~\cref{sec:evaluation}, covering both investigations on the model's stress prediction quality and nonlinear finite element analysis. 

\subsection{Data and PANN model preparation}\label{sec:data_models}

We now provide technical details on the considered datasets and PANN constitutive models.

\subsubsection{Considered homogenization data}\label{sec:data_details}

\begin{figure}[t]
	\centering
    \begin{subfigure}[t]{.26\textwidth}
        \centering
         \includegraphics[width=3.6cm]{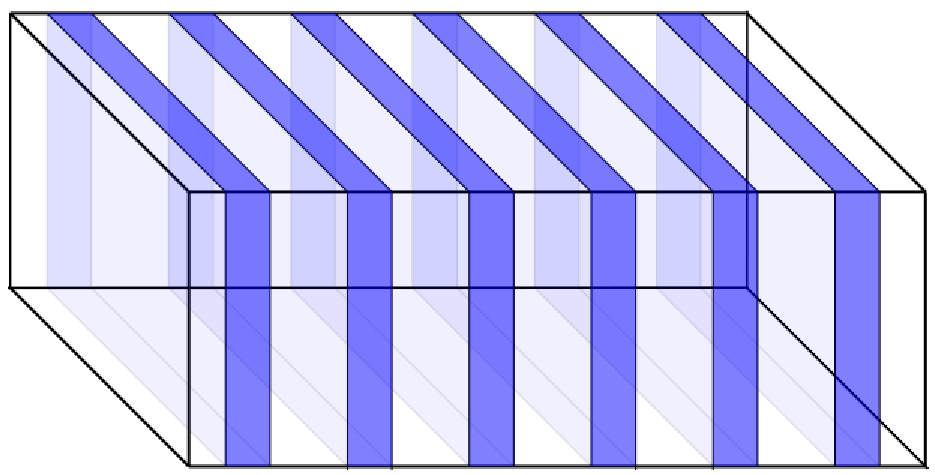}
         \caption{Rank-one laminate (ROL). Image from \parencite{klein2024a}.}
         \label{fig:microstructures:ROL}
    \end{subfigure}%
    ~~~~~
    \begin{subfigure}[t]{.33\textwidth}
        \centering
         \includegraphics[height=1.4cm]{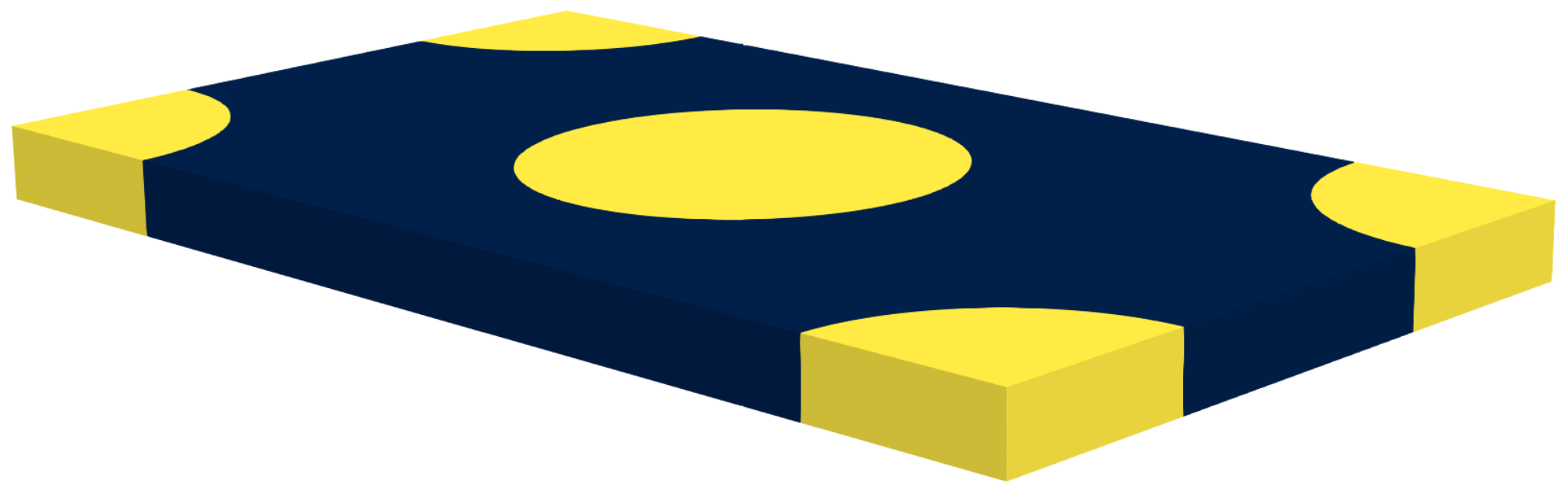} 
         \caption{Matrix with hexagonal fibers (HEX).  Image from \parencite{kalina2025}.}
         \label{fig:microstructures:HEX}
    \end{subfigure}%
    ~~~~~
    \begin{subfigure}[t]{.27\textwidth}
        \centering
         \includegraphics[height=2.8cm]{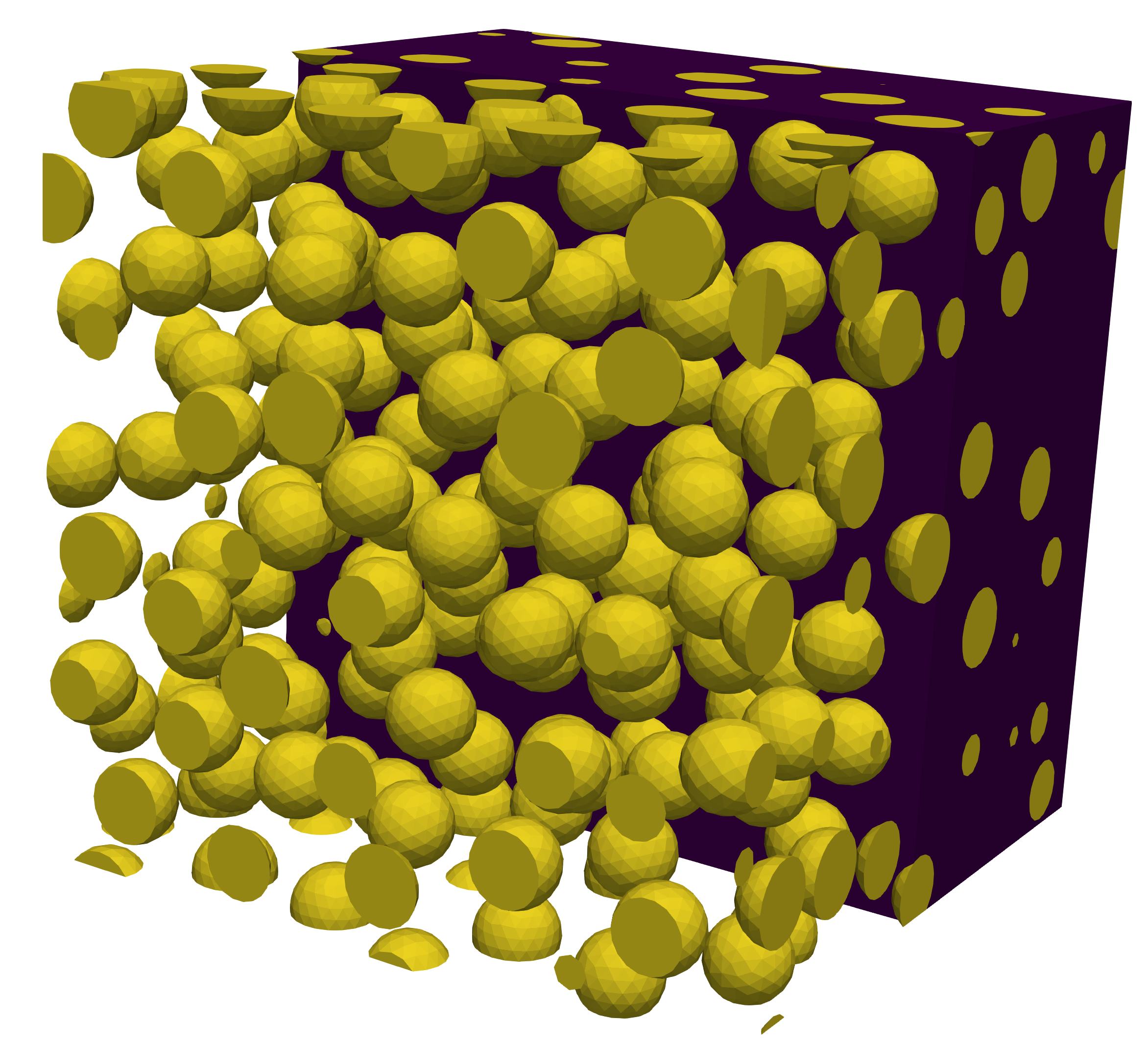}
         \caption{Matrix with random spherical inclusions (RAN).}
         \label{fig:microstructures:RAN}
    \end{subfigure}%
	\caption{Microstructures considered in this work.
	}
	\label{fig:microstructures}
\end{figure}

We consider synthetic homogenization data of three different materials with heterogeneous microstructures, which are visualized in~\cref{fig:microstructures}. The considered composite materials are briefly introduced in the following.

\paragraph{Rank-one laminate (ROL)} The microstructure of a biphasic ROL consists of two phases which are repeatedly stacked, resulting in a transversely isotropic material behavior (Schoenflies group $\cD_{\infty}$), see \cref{fig:microstructures:ROL}. We analytically homogenize the ROL for two different phase contrasts, following the analytical homogenization strategy proposed by \textcite{Debotton_2005}. The material phases are described by the isotropic Mooney-Rivlin model
\begin{equation}\label{eq:micro_behavior_ROL}
\begin{aligned}
\psi_{\circ}=\frac{1}{2}\Big[\mu_{1,\circ}\IoneISO+\mu_{2,\circ}\ItwoISO-2(\mu_{1,\circ}+2\mu_{2,\circ})\log\IthreeISO+\lambda_{\circ}  (\IthreeISO-1)^2\Big]\,,
\end{aligned}
\end{equation}
where $\circ\in\{a,b\}$ assigns the material model to one of the phases. Both phases have equal volume fractions in the laminate.
For phase $a$, the material parameters are given by $\mu_{1,a}=\mu_{2,a}=0.5\text{ MPa}$ and $\lambda_a=50\mu_1^a$. The material parameters of phase $b$ follow from $\mu_{1,b}=\mu_{2,b}=f\,\mu_{1,a}$ and $\lambda_b=f\,\lambda_a$. For the phase contrast $f$, a low value of 1.5 and a high value of 10 are employed. 

For the data generation, we sample the deformation gradient $\bF$ in a wide range of physically admissible deformation modes, using the concentric sampling strategy proposed by \textcite{kunc2019}, see also \textcite[Appendix~B]{klein2022b} for a brief introduction to the method. In the invariant space, the sampled deformation states are within $\IoneISO\in[2.8,\,4.1]$ and $\ItwoISO\in[2.6,\,4.4]$, thus covering a finite deformation regime. From the homogenization, datasets of the form
\begin{equation}\label{eq:dataset_mech}
        \cD=\Big\{\left({^1\bF},\, {^1\bP},\, {^1\bbA}\right),\dotsc,
        \left({^m\bF},\,{^m\bP},\, {^m\bbA}\right)\Big\}\,,
\end{equation}
are obtained containing the deformation gradient $\bF$, the effective first Piola-Kirchhoff stress $\bP$, and the effective tangent $\bbA=\partial_{\bF}\bP$. The overall dataset consists of $m=11\,000$ data points. 

\paragraph{Matrix with hexagonal fibers (HEX)} We consider a soft matrix material with embedded stiff fibers. The fibers are arranged in a hexagonal layout, see \cref{fig:microstructures:HEX}. Thus, the effective material behavior has a hexagonal anisotropy (Schoenflies group $\cD_6$). We consider data from~\textcite{kalina2025}, where a HEX material is numerically homogenized. Therein, both material phases are described by the isotropic Neo-Hooke model
\begin{equation}\label{eq:micro_behavior_FRE}
\begin{aligned}
\psi_{\circ}=\frac{1}{2}\Big[\mu_{\circ}(\IoneISO-\log\IthreeISO)+\frac{\lambda_{\circ}}{2}\big(\IthreeISO-\log\IthreeISO\big)\Big]\quad\text{with}\quad \mu_{\circ}=\frac{E_{\circ}}{2(1+\nu_{\circ})},\,\lambda_{\circ}=\frac{E_{\circ}\nu_{\circ}}{(1+\nu_{\circ})(1-2\nu_{\circ})}\,,
\end{aligned}
\end{equation}
where $\circ$ assigns the material model to either the matrix or the fibre material. The volume fraction of the fibers within the material is 30\%.
For the matrix material, the parameters $E_{\text{matrix}}=1\text{ MPa}$ and $\nu_{\text{matrix}}=0.4$ are employed. The parameters for the fibers follow from $E_{\text{fibre}}=f_E\,E_{\text{matrix}}$ and $\nu_{\text{fibre}}=f_{\nu}\,\nu_{\text{matrix}}$, with the phase contrasts $f_E=10$ and $f_{\nu}=1.1$.

For the data generation, the deformation gradient $\bF$ is sampled in a wide range of physically admissible deformation modes using a method described in~\textcite[Appendix~D]{kalina2025}. In the invariant space, the sampled deformation states are within $\IoneISO\in[2.99,\,4.52]$ and $ \ItwoISO\in[2.9,\,4.9]$, thus covering a finite deformation regime. For this microstructure, we again consider a dataset of the form of~\cref{eq:dataset_mech}, consisting of overall $m=3\,080$ data points.

\begin{remark}\label{rem:hex_ti}
   Note that hexagonal material behavior can, to some extent, be described by transverse isotropy and second order structural tensors \parencite[Sec.~6.2]{Ebbing2010}. In particular, transverse isotropy and hexagonal anisotropy coincide for small deformations around the undeformed configuration \parencite[Sec.~9.3.2]{Haupt2002}. However, the construction of integrity bases for hexagonal anisotropy requires up to sixth order structural tensors \parencite{riemer2025}.     
\end{remark}

\paragraph{Matrix with multiple random spherical inclusions (RAN)} We consider a soft matrix material with multiple spherical inclusions, which are distributed using a random sequential adsorption algorithm \parencite{torquato2002}, see \cref{fig:microstructures:RAN}. All inclusions have the same radius, and the minimum distance between the inclusions is 20\% of their radius. The unit cell contains 300 inclusions which make up 20\% volume fraction. Since the inclusions are distributed randomly, the effective material behavior is approximately isotropic (Schoenflies group $\cK$). We numerically homogenize the RAN material, for which we employ the NGSolve and Netgen libraries \parencite{schoeberl1997,gangl2021}. The geometry is discretized with roughly $6\cdot 10^{5}$ quadratic tetraeder elements. For the microstructure's constitutive behavior, we employ the Ogden model
\begin{equation}\label{eq:ogden}
\psi_{\text{Ogden}}=\sum_{p=1}^{N_0}\frac{\mu_p}{\alpha_p}\left[\sum_{\beta=1}^{3}\left(\bar{\lambda}_{\beta}\right)^{\alpha_p}-3\right]+\frac{\kappa}{4}\left(J^2-2\ln J-1\right)\,,
\end{equation}
with the material parameters 
\begin{equation}
\begin{aligned}
&\text{inclusions:}\,\,\,(\mu_1,\,\alpha_1,\,\kappa)=(1000\text{ MPa},\,2,\,4666.7\text{ MPa})\,,
\\
&\text{matrix:}\,\,\,(\mu_1,\,\mu_2,\,\mu_3,\,\alpha_1,\,\alpha_2,\,\alpha_3,\,\kappa)=(-26.62\text{ MPa},\,29.04\text{ MPa},\,0.0098\text{ MPa},\,-5,\,2.3,\,12,\,800\text{ MPa})\,.
\end{aligned}
\end{equation}
For this choice of material parameters, the matrix and inclusion have a contrast in the initial shear modulus of 10. 

For the data generation, the deformation gradient $\bF$ is sampled in a wide range of physically admissible deformation modes. For this, we employ the method described in~\textcite{kalina2022b}, with the slight adaption that, to account for the effective isotropic material behavior, we consider only main diagonal entries for $\bF$. The sampled deformation states are within $F_{ii}\in[0.8,\,1.3]$ and $J\in[0.9,\,1.15]$, thus covering a finite deformation regime. From the homogenization, datasets of the form
\begin{equation}\label{eq:dataset_mech_2}
        \cD=\Big\{\left({^1\bF},\, {^1\bP}\right),\dotsc,
        \left({^m\bF},\,{^m\bP}\right)\Big\}\,,
\end{equation}
are obtained containing the deformation gradient $\bF$ and the effective first Piola-Kirchhoff stress $\bP$. We overall prescribe $1\,000$ deformation gradients, out of which 804 converge. Thus, the overall dataset consists of $m=804$ data points. 

\paragraph{Ellipticity of the homogenization data}
Polyconvex constitutive models can reasonably represent only elliptic material behavior. Thus, we want to consider the effective material behavior of the metamaterials only within their elliptic regime. It is important to note that even if the microstructure of the material is described by polyconvex (and thus elliptic) material models, the homogenized behavior might still be non-elliptic \parencite{Rudykh_deBotton_2012,KBertoldi_11_01}. To assess ellipticity of the homogenization datasets, the effective tangent $\bbA$ is required, which we have for the ROL and HEX materials. For the ROL material with low phase contrast, all datapoints are elliptic, whereas for the ROL material with high phase contrast, approximately 25\% of the datapoints are non-elliptic. The loss of ellipticity occurs for datapoints with large mechanical loads. Thus, the data considered for the ROL material with large phase contrast is at the verge of loss of ellipticity. For the HEX material, approximately 3\% of the datapoints are non-elliptic. For all composite materials, we exclude the non-elliptic datapoints in the further investigations. While for the RAN material, we do not homogenize the tangent $\bbA$, the excellent performance of the signed singular value-based polyconvex PANN model for this dataset suggests that this material behaves elliptic within the considered deformation range, cf.~\cref{sec:stress_pred}. 

\subsubsection{Considered PANN models}\label{sec:micro_PANNs}

For all anisotropic composite materials considered in this work, the preferred directions are known and the structural tensors are chosen accordingly. The hyperparameters are chosen based on preliminary studies and in accordance to literature standards \parencite{kalina2025,klein2022b}. In addition, we perform hyperparameter studies in~\cref{sec:application_micro_calib} to investigate some model properties. We employ the following PANN models, which are introduced in detail in~\cref{sec:PANNs,sec:PANNsSSV}:

\paragraph{PANN--$\cI/\cI^*$ models using structural tensor-based invariants} We employ both polyconvex and non-polyconvex PANN models using structural tensor-based invariants for isotropy and transverse isotropy. In the following, such models that are polyconvex are denoted by PANN--$\cI$, and non-polyconvex models by PANN--$\cI^*$. We employ the invariants (cf.~\cref{eq:invs_iso,eq:invs_ti_pc})
\begin{equation}\label{eq:iso_invar_set}
\begin{aligned}
&\bcI^{\text{iso}} = 
\left\{\begin{aligned}
&\big(\IoneISO,\,\ItwoISO,\,J\big)&&\in\bbR^{3}\,, && (\text{PANN--}\cI^*)\\
&\big(\IoneISO,\,\ItwoISO,\,J,\,-J
\big)&&\in\bbR^{4}\,, && (\text{PANN--}\cI)\\
\end{aligned}\right. \,,
\\
&\bcI^{\text{ti}} = 
\left\{\begin{aligned}
&\big(\IoneISO,\,\ItwoISO,\,J,\,\IoneTI,\,\ItwoTI\big)&&\in\bbR^{5}\,, && (\text{PANN--}\cI^*)\\
&\big(\IoneISO,\,\ItwoISO,\,J,\,-J,\,\IoneTI,\,\ItwoTI,\,\IoneISO-\ItwoTI,\,\ItwoISO-\ItwoTI\big)&&\in\bbR^{8}\,, && (\text{PANN--}\cI)\\
\end{aligned}\right. \,.
\end{aligned}
\end{equation} 
A comprehensive motivation for these choices of invariant sets is provided in~\textcite[Secs.~3.3.1 and 3.3.2]{klein2026a}. For the NN potentials, FFNNs with a single hidden layer are employed ($H=1$), with $n_1=16$ nodes for the ROL and HEX materials and $n_1=32$ nodes for the RAN material. These architectures offer sufficient flexibility while maintaining a moderate overall network size. 
 
\paragraph{PANN--$\nu$ model using signed singular values} We employ the signed singular value-based polyconvex PANN model for isotropy. For the NN potential, we employ FFNNs with again only a single hidden layer ($H=1$) and $n_1=32$ nodes. In the following, this model is denoted by PANN--$\nu$.

\subsubsection{Model calibration}\label{sec:application_micro_calib}

We now introduce details on the model calibration. Recall that the weights and biases of the NNs form the set of parameters $\bcP$ of the PANN model. 

\paragraph{Formulation of the loss function} The overall dataset is split into a calibration dataset consisting of $m_{\text{c}}$ datapoints and a test dataset consisting of $m_{\text{t}}=m-m_{\text{c}}$ datapoints. We employ 800 random datapoints for calibration for the ROL and HEX materials and 400 for the RAN material. To fit the PANN model parameters $\bcP$ to a given dataset $\cD$, a loss function $\mathfrak{L}$ is minimized, for which we consider the  the mean squared error (MSE) of the stress predictions, i.e.,
\begin{equation}\label{eq:stress_loss_micro_P}
\begin{aligned}
        \mathfrak{L}(\bcP)=\frac{1}{m_{\text{c}}}\sum_{i=1}^{m_{\text{c}}}\frac{1}{9\star}\left\|{^i\bP}-{^i\bP}^{\text{model}}({^i\bF},\,\bcP)\right\|^2\,,
\end{aligned}
\end{equation}
where $\star$ is introduced to make the loss function dimensionless. We set $\star=\text{MPa}^2$ for the ROL and RAN materials and $\star=\text{kPa}^2$ for the HEX material. All PANN models are solely calibrated on the stresses, i.e., the first derivatives of the strain energy, even if data for the second derivative is available. The calibration of the PANN models through their gradients is referred to as Sobolev training \parencite{Czarnecki2017SobolevTF,vlassis2021}. Since in this work, we focus on the prediction of the derivatives of the energy potentials, we do not consider energy values in the calibration process. This decision is supported by the observation that including energy data in the calibration process does not enhance the prediction quality of the PANN model's derivatives \parencite{klein2022a}. 

\paragraph{Implementation and calibration details} The transversely isotropic PANN models are implemented and calibrated in TensorFlow 2.10.0, while the isotropic PANN models are implemented and calibrated in KLAX (\url{https://drenderer.github.io/klax/}). The parameter optimization is performed with the ADAM optimizer, employing the full calibration dataset, a batch size of 32, without loss weighting, and with a learning rate of $5 \cdot 10^{-3}$. For the ROL and HEX materials, the PANN models are calibrated for $5\cdot 10^3$ epochs, while for the RAN materials, the PANN models are calibrated for  $5\cdot 10^5$ steps. Note that the definition of calibration epochs and steps varies between TensorFlow and KLAX. In each investigation, the PANN models are calibrated five times, and the model with the lowest test loss is used for evaluation. By that, we account for the random initialization of the NN parameters and stochastic effects in the optimization process. The growth and normalization terms are employed as described in~\textcite{klein2026a} and \textcite{Geuken_Kurzeja_Wiedemann_Mosler_2025}.

\subsection{NN hyperparameter study}\label{sec:application_micro_calib}

\begin{figure}[t!]
\centering
\includegraphics[width=0.7\textwidth]{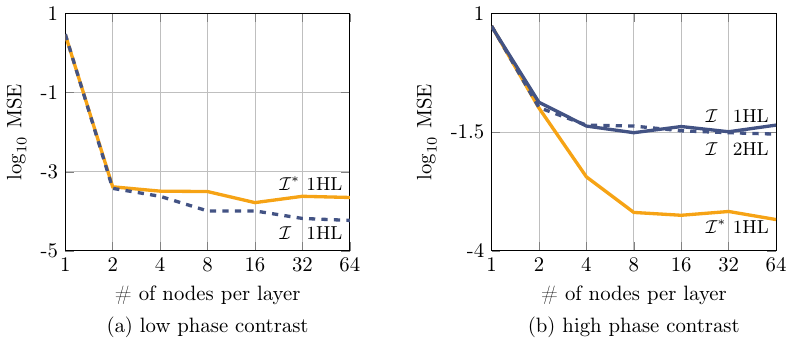}
\caption{Hyperparameter study for PANN models applied to the ROL data. Calibration MSE for different number of nodes and hidden layers (HL). For each setting, five PANN models are calibrated, and the results for the model instance with the lowest calibration MSE are visualized.
}  
\label{fig:hyperparam_ROL}
\end{figure}

To demonstrate that once a NN architecture is sufficiently large for the targeted application, there is no benefit in further increasing its size, we perform an initial hyperparameter study. For this, we consider the PANN--$\cI/\cI^*$ models applied to the ROL material with low and high phase contrast. Specifically, we examine the following three PANN variants: a polyconvex PANN--$\cI$ model with a single hidden layer ($H=1$), a polyconvex PANN--$\cI$ model with two hidden layers ($H=2$), and a non-polyconvex PANN--$\cI^*$ model with a single hidden layer ($H=1$). The number of nodes in each hidden layer $n_h$ is varied between one and 64. Each PANN model is calibrated five times using the calibration parameters as for the ROL study in~\cref{sec:stress_pred}. The MSE results are visualized in~\cref{fig:hyperparam_ROL}.

For the ROL material with low phase contrast, both PANN approaches achieve a comparable performance, where NN architectures with as few as a single hidden layer and eight nodes are sufficiently flexible, and further increasing the network size does not considerably improve the performance of the PANN models. A similar observation holds for the PANN--$\cI^*$ model applied to the ROL material with high phase contrast. In contrast, across all tested architectures, the PANN--$\cI$ model consistently exhibits higher MSE values than its non-polyconvex counterpart when applied to the ROL material with high phase contrast. For example, a PANN--$\cI$ model with two hidden layers and 32 nodes in each layer has approximately 20 times as many trainable parameters as the one with a single hidden layer with 16 nodes. Still, both models show a very similar performance. 

To anticipate the following investigations, when a polyconvex PANN model shows a reduced performance, this is not caused by an insufficient number of model parameters, but rather due to the inherent constraints imposed by polyconvex constitutive modeling. Together with the observation that PANN models are robust against overfitting \parencite{linden2023}, thereby mitigating potential drawbacks typically associated with overparameterization, the findings of this hyperparameter study further justify the pragmatic selection of NN architectures that we employ. Once a NN is sufficiently large for the targeted application, further increasing its size yields no notable advantage and only increases training and inference duration and cost. 

\begin{figure}[t!]
\centering
\includegraphics[width=\textwidth]{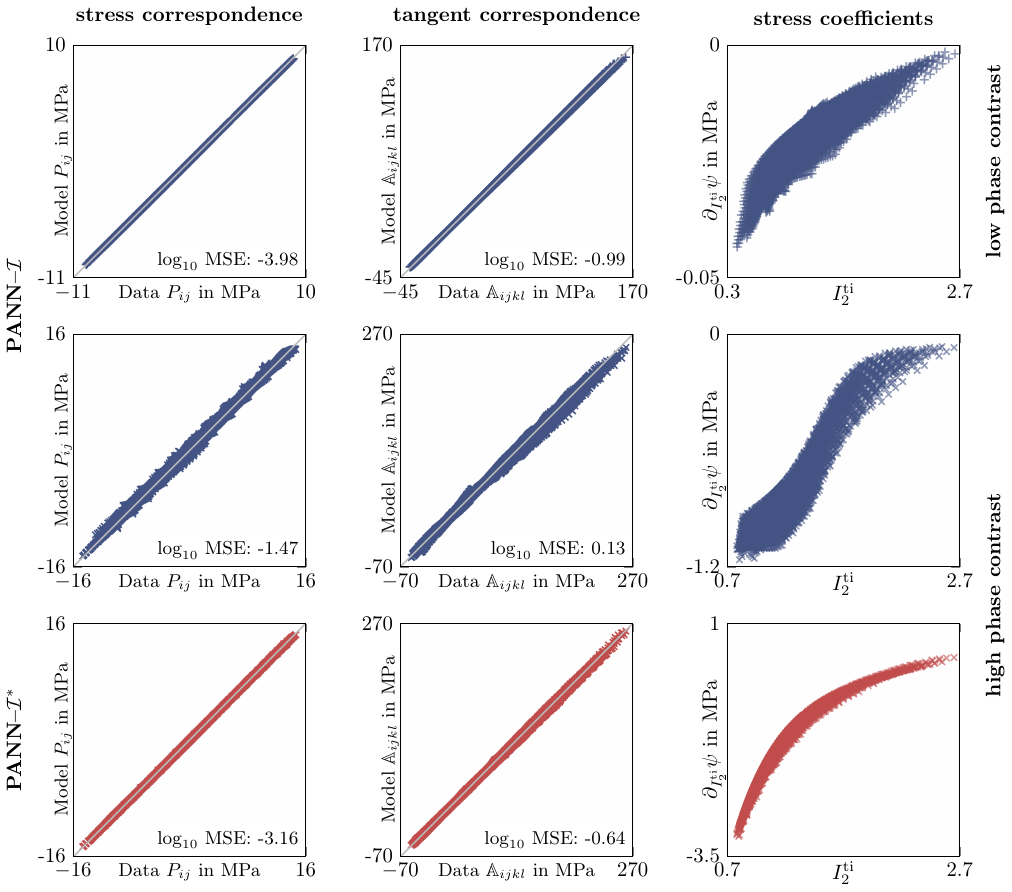}
\caption{Performance of different PANN models for the ROL data.} 
\label{fig:ROL}
\end{figure}

\subsection{PANN model evaluation}\label{sec:evaluation}

In the following, we investigate the performance of different PANN constitutive models for the homogenization data. 
We investigate the stress (and, for the ROL materials, also the tangent) prediction quality of the models in~\cref{sec:stress_pred}, followed by nonlinear finite element analysis with the PANN models in~\cref{sec:FEA}.


\subsubsection{Stress prediction}\label{sec:stress_pred}

\paragraph{Application to the ROL data} We apply transversely isotropic PANN--$\cI$ and PANN--$\cI^*$ models to the ROL data with both a low and a high phase contrast. The models are calibrated solely on the stress values, and in~\cref{fig:ROL}, correspondence plots are provided for predictions of both the stress and the tangent for the test dataset. 

For the ROL with a low phase contrast, the PANN--$\cI$ model shows an excellent performance (see the top row in \cref{fig:ROL}). Notably, although only calibrated on the stress values, it also has an excellent prediction of the tangent values. In contrast, for the ROL with high phase contrast, the PANN--$\cI$ model exhibits significant deviations from the true material behavior, affecting both stress and tangent predictions (middle row). Conversely, the non-polyconvex PANN--$\cI^*$ model achieves an excellent performance for this material (bottom row). In particular, as it learns to accurately predict the tangent, it also adopts the ellipticity of the ROL data, and it is numerically investigated that the model is elliptic for all deformation modes contained in the test dataset. This demonstrates one of the mitigation strategies discussed in~\cref{sec:mitigation}: 
\begin{center}
\fbox{\parbox{0.98\linewidth}{
Given suitable calibration data, \textbf{non-polyconvex model formulations} can be employed.}}
\end{center}

The different performances of polyconvex and non-polyconvex PANN models can be better understood by examining the stress coefficients, i.e., the derivatives of the PANN models w.r.t.\ the strain invariants, also visualized in~\cref{fig:ROL} in the right column. For non-polyconvex PANN models, the stress coefficients can basically adopt arbitrary functional forms. In contrast, the polyconvex PANN models are subject to monotonicity and convexity constraints (cf.~\cref{sec:invs}). As we discuss in~\textcite[Sec.~3.3.1]{klein2026a}, these constraints set by polyconvexity can be partially relaxed by extending the polyconvex integrity basis with suitable invariants, which enables the polyconvex models to represent negative stress coefficients to a certain extent.
For the ROL with low phase contrast, the stress coefficients take only slightly negative values, which can still be captured by the PANN--$\cI$ model. However, for the ROL with high phase contrast, the PANN--$\cI^*$ model shows more pronounced negative stress coefficients, which apparently are required for an accurate representation of the material behavior. Owing to the inherent constraints of the PANN--$\cI$ model, it cannot fully reproduce these negative stress coefficients and therefore exhibits a reduced performance in modeling the ROL with high phase contrast. Thus:
\begin{center}
\fbox{\parbox{0.98\linewidth}{
The \textbf{sufficient but not necessary} monotonicity and convexity constraints imposed on polyconvex potentials using structural tensor-based invariants can be overly restrictive.
}}
\end{center}
As demonstrated in the hyperparameter study in~\cref{fig:hyperparam_ROL}, increasing the width and depth of the NN architectures does not enhance the performance of the PANN--$\cI$ model applied to the ROL data with high phase contrast. This supports the claim that the limited flexibility of this model stems from the constraints imposed by polyconvexity, rather than an insufficient number of model parameters. We reported similar observations for polyconvex PANN constitutive models applied to homogenization data of electro-elastic microstructured materials in \textcite{klein2024a} and experimental data of soft rubber-like materials in \textcite{klein2025}.

\begin{figure}[t!]
\centering
\includegraphics[width=0.6667\textwidth]{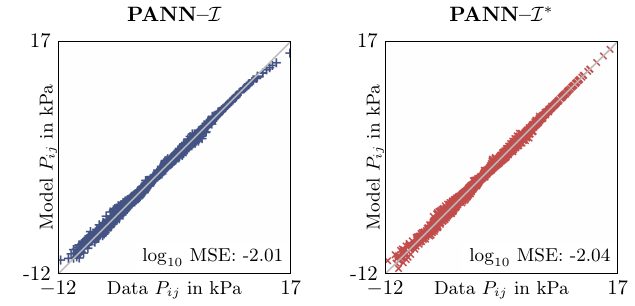}
\caption{Performance of different PANN models for the HEX data.} 
\label{fig:HEX}
\end{figure}

\paragraph{Application to the HEX data} We apply transversely isotropic PANN--$\cI$ and PANN--$\cI^*$ models to the HEX data. In~\cref{fig:HEX}, correspondence plots are provided for the prediction of the stress for the test dataset. Both models show a very similar behavior and achieve only moderate accuracy, which can be explained as follows. The ground truth material behavior has a hexagonal anisotropy. While hexagonal material behavior can, to some extent, be described by transverse isotropy and second order structural tensors (cf.~Remark~\ref{rem:hex_ti}), the construction of integrity bases for hexagonal anisotropy requires up to sixth order structural tensors \parencite{riemer2025}. To the best of our knowledge, no polyconvex invariants based on sixth-order structural tensors have been proposed. Thus:
\begin{center}
\fbox{\parbox{0.98\linewidth}{
For some anisotropies, \textbf{no polyconvex integrity or functional bases} are available, and polyconvex constitutive models rely on incomplete inaviant sets. This leads to a loss of information and limits the flexibility of the models.
}}
\end{center}
For the HEX material, this claim is supported by the moderate performance of the non-polyconvex model, for which we also employ the transversely isotropic integrity basis. In particular, the moderate performance of the PANN--$\cI^*$ model indicates that this is not caused by the imposed monotonicity and convexity constraints on the potential, but rather by the incomplete set of invariants. For the non-polyconvex model, the non-polyconvex integrity basis for hexagonal anisotropy could be employed \parencite{riemer2025} instead. In this case, the PANN--$\cI^*$ model has been shown to have achieve an excellent performance for the HEX data \parencite{kalina2025}. 

\begin{figure}[t!]
\centering
\includegraphics[width=\textwidth]{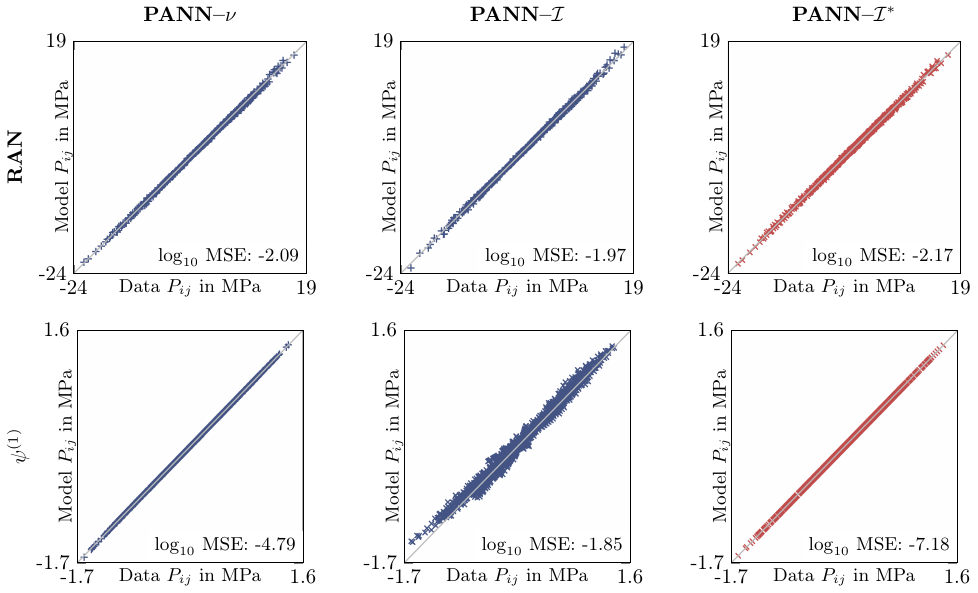}
\caption{Performance of different PANN models for the RAN data and for data generated with the analytical potential $\psi^{(1)}$ in~\cref{eq:ana_pot_ex_1}.} 
\label{fig:RAN}
\end{figure}

\paragraph{Application to the RAN data and to an analytical potential} We apply isotropic PANN--$\cI$, PANN--$\cI^*$, and PANN--$\nu$ models to the RAN material. In~\cref{fig:RAN}, correspondence plots are provided for the stress predictions for the test dataset. All PANN models perform very similar and show an excellent performance in predicting the ground truth stress behavior. 

Next, we investigate the performance of the PANN models for data generated with the analytical Mooney-Rivlin-type potential from \cref{eq:ana_pot_ex_1} with now $ \mu_1=1\text{ MPa},\,\mu_2=3\text{ MPa}$. As we discuss in~\cref{sec:hessian}, this potential does not fulfill the sufficient conditions of polyconvex invariant-based modeling. Our investigation thus deliberately challenges the polyconvex PANN--$\cI$ model. All calibration details are chosen as for the RAN material, and in~\cref{fig:RAN}, correspondence plots are provided for the stress of the test dataset. The PANN--$\cI$ model shows only a moderate performance in predicting the ground truth material behavior. Despite the flexibility of the NN potential, the model cannot accurately represent this material behavior which does not fulfill the sufficient conditions for polyconvexity of invariant-based potentials. In contrast, both the PANN--$\cI^*$ and PANN--$\nu$ models show an excellent performance for this dataset. This demonstrates one of the mitigation strategies discussed in~\cref{sec:mitigation}:
\begin{center}
\fbox{\parbox{0.98\linewidth}{
When one specific polyconvex formulation fails to accurately represent the considered material behavior, a \textbf{different polyconvex formulation} might perform better.}}
\end{center}

\paragraph{Discussion} The homogenized behavior of microstructured materials sets high challenges to polyconvex constitutive models. While, in some of our investigations, polyconvex PANN models show an excellent performance, in other cases, they seem to be overly restrictive.\footnote{A limited performance of some polyconvex PANN models was also reported for experimental data of soft rubber-like materials \parencite{klein2025,wollner2026}.} This can be explained as follows. It is well-known that \textbf{homogenized microstructures are prone to a loss of ellipticity} \parencite{Rudykh_deBotton_2012,Khajehtourian_Kochmann_2021}.
Consequently, even though we only considered the homogenized material behavior within its elliptic regime, many of the data points are at the verge of a loss of ellipticity. Indeed, the aim of polyconvex constitutive modeling is to represent material behavior that is elliptic. However, polyconvexity is sufficient but not necessary for ellipticity (cf.~\cref{sec:inv_conv_examples}). Thus, even when a polyconvex model with universal approximation properties for the considered material symmetry would be available, it might not be able to adequately represent the elliptic material behavior at hand. Moreover, the way that polyconvex anisotropic constitutive models are formulated contains even more sufficient but not necessary conditions. For instance, formulating polyconvex material models using structural tensor-based invariants as convex and monotonic is sufficient but not necessary to ensure polyconvexity (cf.~\cref{sec:invs}). Overall, this results in a cascade of restrictions on the function space that the polyconvex constitutive models can represent, in particular for the often very challenging anisotropies of homogenized microstructures. 
Furthermore, as the phase contrast in fibre-reinforced or laminate structures increases, their homogenized response tends to lose ellipticity at progressively smaller deformations \parencite{Rudykh_deBotton_2012}. This aligns with our observation that the polyconvex PANN models generally perform well for moderate phase contrasts, but tend to fail for larger phase contrasts. 


\begin{figure}[t!]
	\centering
	\begin{tabular}{>{\centering\arraybackslash}p{4.5cm}@{\hspace{1.5cm}} >{\centering\arraybackslash}p{4.5cm}@{\hspace{1.5cm}} >{\centering\arraybackslash}p{4.5cm}}
		\includegraphics[height=4cm]{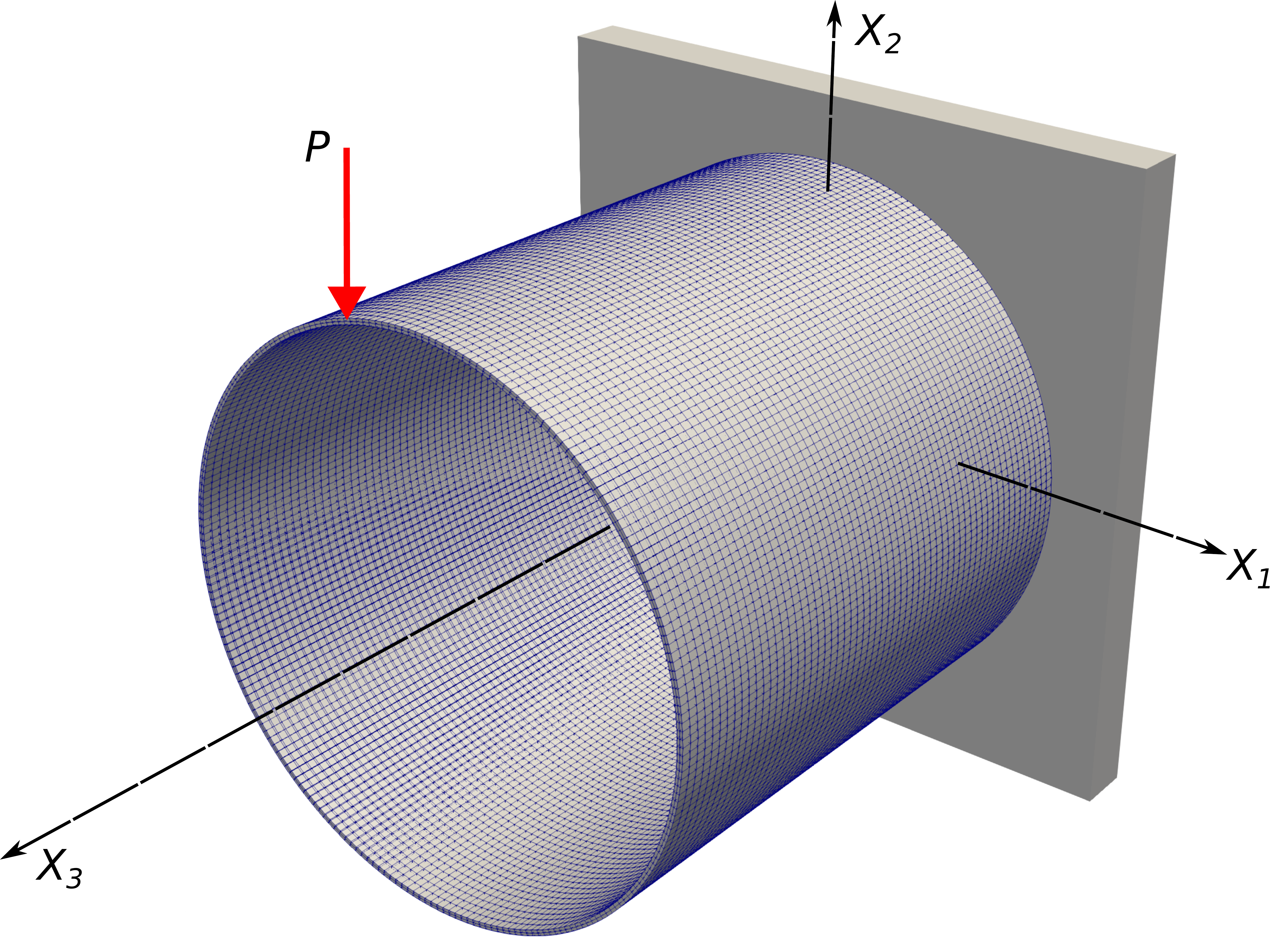} &

		\includegraphics[height=4.45cm]{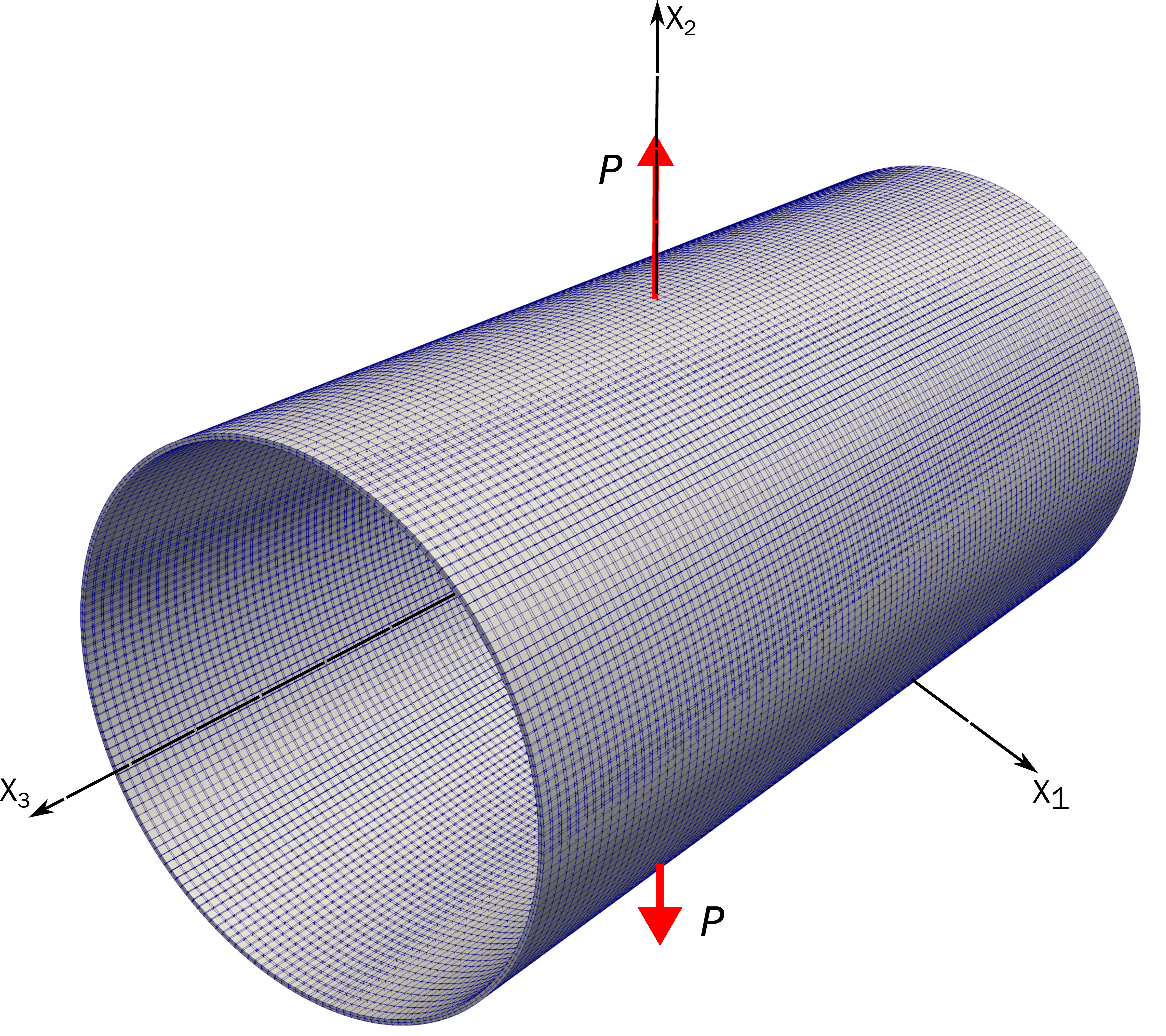}&
        \includegraphics[height=4.45cm]{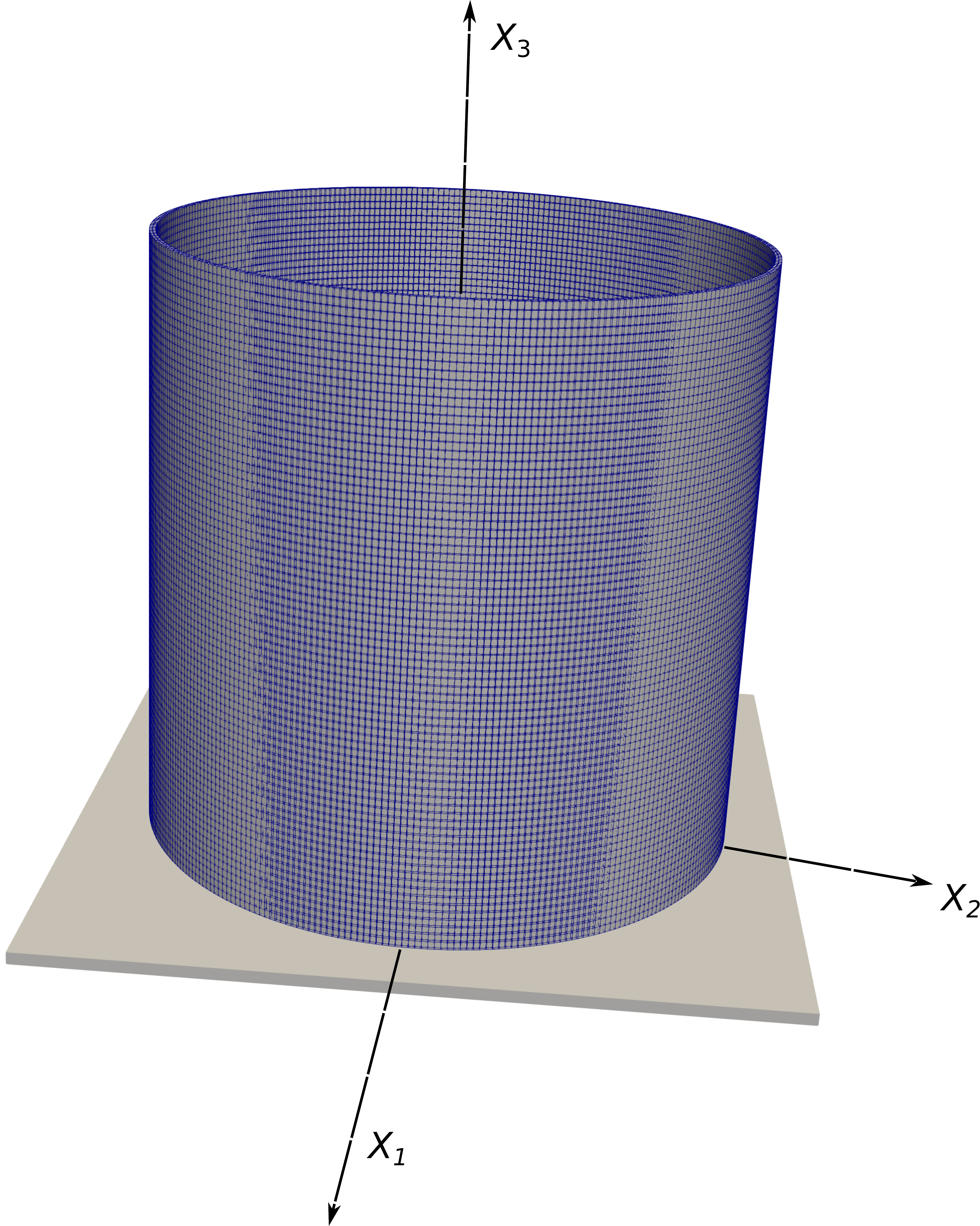}
        \\
		{\small\textbf{(a)}} & {\small \textbf{(b)}} & {\small\textbf{(c)}}
	\end{tabular}
	\caption{Considered simulation scenarios for the FEA.
    }
	\label{fig:sim_scenarios}
\end{figure}

\begin{figure}[hp!]
\centering
\includegraphics[width=\textwidth]{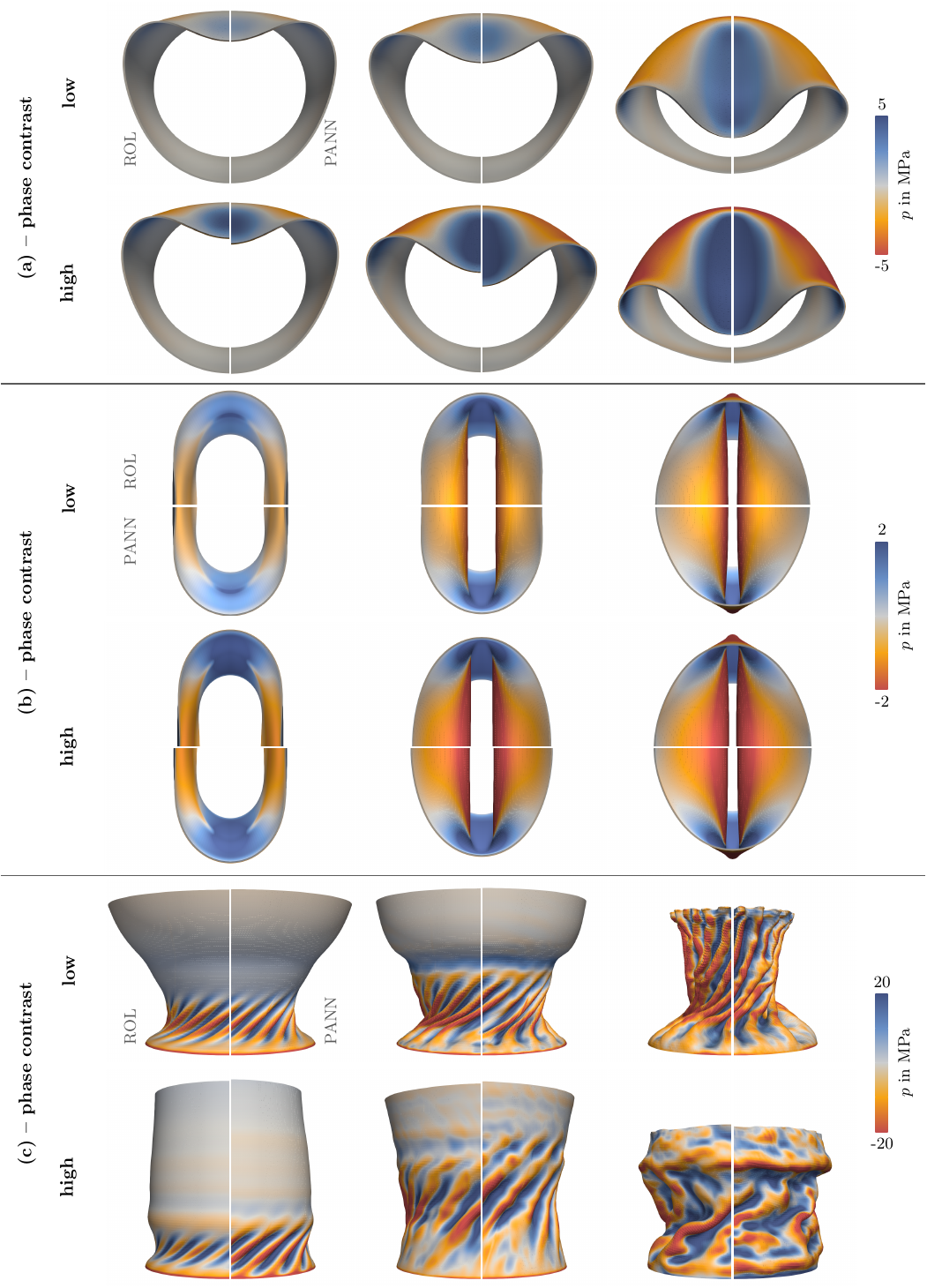}
\caption{Results for the simulation scenarios (a--c) with a ROL material with low / high phase contrast and a corresponding PANN--$\cI$ constitutive model. Scaling factor of one, contour plot distribution for the hydrostatic pressure $p$. Visualization for different load factors: (a): [0.3,0.44,1.0], (b): [0.07,0.3,1.0], (c) low: [0.62,0.71,1.0], (c) high: [0.51,0.7,1.0].
} 
\label{fig:paraview}
\end{figure}

\subsubsection{Nonlinear finite element analysis}\label{sec:FEA}

In the previous section, we investigated the stress prediction quality of different PANN constitutive models calibrated to homogenization data of microstructured materials. Now, we investigate their performance in nonlinear finite element analysis (FEA), and discuss some further open questions. 

\paragraph{Simulation details} We employ the ROL material with a low and a high phase contrast and corresponding PANN models using structural tensor-based invariants. For the PANN models, we employ the best calibrated model instance from the previous section. For the simulations with a low phase contrast, we consider only a polyconvex PANN--$\cI$ constitutive model, while for the simulations with a high phase contrast, we consider both polyconvex PANN--$\cI$ and non-polyconvex PANN--$\cI^*$ constitutive models. For the FEA, both the analytical ROL homogenization and the PANN models are implemented in the Julia-based package Gridap.jl \parencite{gridap1,gridap2}, in combination with the package HyperFEM (https://github.com/MultiSimOLab/HyperFEM.jl) \parencite{DanielAdvancedMaterials}. We investigate three simulation scenarios of hollow cylinders, which are outlined in~\cref{fig:sim_scenarios}. In all cases, self contact is not considered. The geometries are discretized with tri-quadratic (Q2) elements, and mesh convergence was ensured in preliminary studies. The remaining simulation details are as follows: \textbf{(a)} The cylinder has an inner / outer radius of 0.049 / 0.05 m and a length of 0.1 m. The mesh consists of $240\times 60 \times 2$ elements in the circumferential, longitudinal, and radial directions, respectively. The cylinder is completely clamped at $x_3=0$, while a nodal load is applied on the opposite end. For the ROL with low and high phase contrast, a nodal load of of $P=3\cdot 10^{-6}$ N and $P=1.2\cdot 10^{-5}$ N is applied, respectively. \textbf{(b)} The cylinder has an inner / outer radius of 0.049 / 0.05 m and a length of 0.2 m. The mesh consists of $200\times 100 \times 2$ elements in the circumferential, longitudinal, and radial directions, respectively. The cylinder gets pulled by two opposite forces. For the ROL with low and high phase contrast, a nodal load of of $P=6.4\cdot 10^{-6}$ N and $P=1.92\cdot 10^{-5}$ N is applied, respectively. \textbf{(c)} The cylinder has an inner / outer radius of 0.49 / 0.5 m and a length of 1 m. The mesh consists of $80\times 80 \times 2$ elements in the circumferential, longitudinal, and radial directions, respectively. Unlike the previous examples, which were subject to static loading, this scenarios involves an initial angular velocity of $\Omega_0=1000\,\text{rad}/\text{s}$ around the $x_3$ axis. The cylinder has a density of $\rho_0=10^{-5}\,\text{kg/m}^3$ and is completely clamped at $x_3=0$.

\begin{figure}[t!]
\centering
\includegraphics[width=\textwidth]{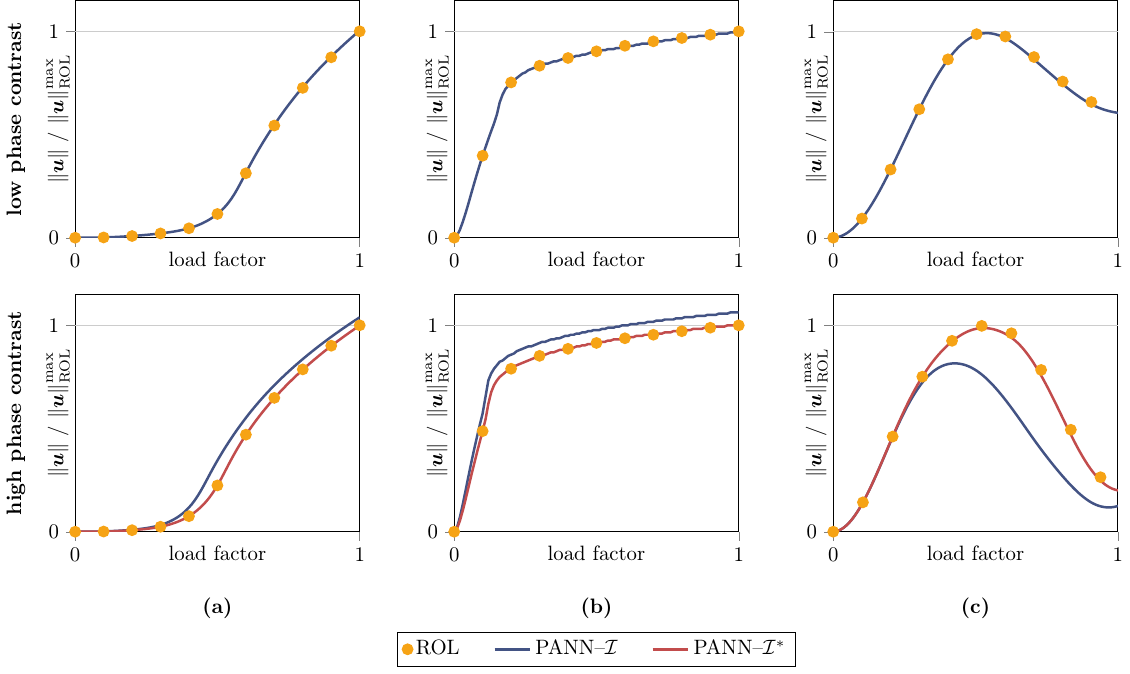}
\caption{Displacement norm for PANN models compared to the ground truth ROL material behavior for the different simulation scenarios. The values are normalized by the maximum displacement norm of the ROL simulation in each scenario.
} 
\label{fig:def_norm}
\end{figure}

\paragraph{When it comes to the stress prediction quality of a constitutive model -- how good is good enough?} This question arises as, in engineering practice, our interest extends beyond stress-strain curves. Instead, we employ constitutive models within simulation scenarios to predict the behavior of engineering components made from the material under investigation. In~\cref{fig:paraview}, the simulation results obtained with the polyconvex PANN--$\cI$ model are compared to those of the ground truth ROL material for various load factors. While for a low phase contrast, both seem to perfectly coincide, for a high phase contrast, visible deviations can be observed. A more quantitative investigation is provided in~\cref{fig:def_norm}, where the displacement norms of the simulations obtained with the ROL and PANN constitutive models are compared. Again, for a low phase contrast, the polyconvex PANN model shows an excellent agreement with the ROL material, while for a high phase contrast, visible deviations occur. These findings are consistent with the observations in~\cref{sec:stress_pred}, where the PANN--$\cI$ model performed excellently for a low phase contrast but only moderately for a high phase contrast (see \cref{fig:ROL}). Nevertheless, even for a high phase contrast, the deviations between the PANN- and ROL-based simulations remain mostly moderate. This suggests that the predicted stresses and tangents of the PANN--$\cI$ model could still be sufficiently accurate for at least some simulation scenarios. Clearly, in other scenarios where errors might accumulate, the prediction quality of the PANN--$\cI$ models for the ROL with a high phase contrast could be inaccurate \parencite[Sec.~5.1]{klein2024a}. This underlines that the quality of a constitutive model cannot be assessed solely based on correspondence plots or loss metrics. Rather, its adequacy must be evaluated in the context of the intended application. 

\vspace{-0.2cm}

\paragraph{Can a non-polyconvex PANN constitutive model be applied in complex simulation scenarios?} This question arises as non-polyconvex constitutive models are not elliptic by construction, and ellipticity ensures numerically stable simulations (cf.~\cref{sec:eng_perspective}) \parencite{Schroeder_Neff_Balzani_2005}. The simulation results obtained with the non-polyconvex PANN--$\cI^*$ model and the ROL material with a high phase contrast are in excellent agreement (see \cref{fig:def_norm}). Remarkably, the PANN--$\cI^*$ model enables stable numerical simulations even in the presence of numerically challenging phenomena such as large deformations and buckling (see \cref{fig:paraview}). Although the PANN--$\cI^*$ model is not elliptic by construction, it has adopted the ellipticity of the ground truth data in the calibration process (see \cref{sec:stress_pred}), which results in numerical stability of the considered simulation scenarios. This can be explained as follows. When calibrating PANN constitutive models to some material behavior, the model can learn to represent not only the stress $\bP$, but also the tangent $d_{\bF}\bP$, in some cases, even when the model is calibrated solely on stress data. Since ellipticity is directly linked to the tangent, the PANN--$\cI^*$ model can adopt the ellipticity of the calibration data. However, this is only possible if sufficient elliptic calibration data is available, and ellipticity might still be lost outside the calibration regime.

\vspace{-0.1cm}

\section{Conclusion}\label{sec:conc}

What is the intended application of polyconvex constitutive models? In this work, we adopt the perspective that they are employed to represent elliptic material behavior, in which case we would like the models to perfectly represent the ground truth material behavior. From this, we derived the {limitations} of polyconvex constitutive modeling: In some cases, polyconvex constitutive models fail to adequately represent material behavior although it is elliptical. We summarized the various theoretical reasons for the limited applicability of polyconvex constitutive models. In a nutshell, the limitations of polyconvex constitutive models arise due to a cascade of \emph{sufficient but not necessary} conditions: Polyconvexity is \emph{sufficient but not necessary} for ellipticity. Moreover, the way that polyconvex constitutive models are constructed usually entails \emph{sufficient but not necessary} conditions. First of all, only polyconvex input quantities are used as inputs of the energy potentials -- however, for many anisotropies, no polyconvex integrity or functional bases are available. Furthermore, we employ \emph{sufficient but not necessary} monotonicity and convexity constraints on the energy potential. Simply put, not all elliptic functions can be represented by polyconvex functions, and in most cases, not all polyconvex functions can be represented by the specific polyconvex constitutive model formulation one might employ.

\medskip

We investigated the practical limitations of polyconvex constitutive models at the example of physics-augmented neural network (PANN) constitutive models. For this, we considered polyconvex PANN models using structural tensor-based invariants and models using signed singular values. We investigated the performance of the PANN constitutive models by applying them to different datasets of homogenized microstructured materials. While for some material behaviors, the polyconvex PANN models showed an excellent performance, in other scenarios, they showed only a moderate performance. We discussed different {mitigation} strategies that can be applied when polyconvex constitutive models perform poorly. The benefits, limitations, and mitigation strategies can be summarized as follows:

 \begin{figure}[h!]
        \centering
\includegraphics[width=0.9846\textwidth]{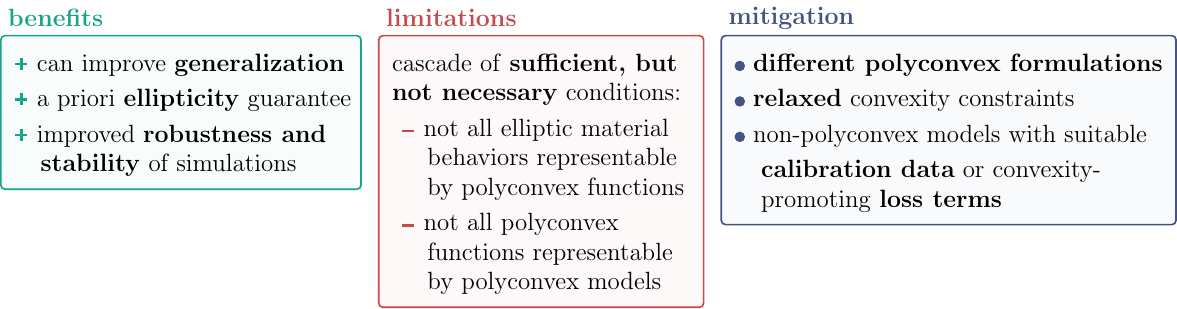}
 \end{figure}

\vspace{-0.2cm}

\noindent
The inductive biases inherent to polyconvex constitutive modeling can substantially improve the performance of a model, but they can also limit its applicability. Ultimately, constitutive modeling remains a trade-off between structure and flexibility.

\newpage

\noindent
\textbf{CRediT authorship contribution statement.} 
\textbf{D.K.\ Klein:} Conceptualization, Methodology, Software, Formal analysis, Investigation, Writing -- original draft, Writing -- review and editing, Visualization.
\textbf{R. Ortigosa:} Conceptualization, Software, Investigation, Funding acquisition, Writing -- review and editing. 
\textbf{H.T.\ Roth:} Software, Investigation, Writing -- review and editing.  
\textbf{K.A.\ Kalina:} Methodology, Writing -- review and editing. 
\textbf{J. Mart\'inez-Frutos:} Funding acquisition, Conceptualization, Writing -- review and editing. 
\textbf{M. Kästner:} Funding acquisition, Writing -- review and editing. 
\textbf{O. Weeger:} Conceptualization, Funding acquisition, Writing -- review and editing. 
\vspace*{1ex}

\noindent
\textbf{Conflict of interest.} The authors declare that they have no conflict of interest.
\vspace*{1ex}

\noindent
\textbf{Acknowledgment.} 
D.K.\ Klein and O.\ Weeger acknowledge the financial support provided by the Deutsche Forschungsgemeinschaft (DFG, German Research Foundation, project number 492770117) and the Graduate School of Computational Engineering at TU Darmstadt. H.T.\ Roth, K.A.\ Kalina and M.\ Kästner want to thank the DFG for the financial support within the Research Training Group
GRK 2868 D${}^3$–Project Number 493401063. R. Ortigosa and J. Mart\'inez-Frutos acknowledge the support of grant PID2022-141957OA-C22 funded by MICIU/AEI/10.13039/501100011033 and by ``ERDF A way of making Europe''. They also acknowledge the support provided by the Autonomous Community of the Region of Murcia, Spain through the programme for the development of scientific and technical research by competitive groups (21996/PI/22), included in the Regional Program for the Promotion of Scientific and Technical Research of Fundacion Seneca - Agencia de Ciencia y Tecnologia de la Region de Murcia
\vspace*{1ex}

\noindent
\textbf{Data availability.}
After acceptance of the manuscript, the authors provide access to the homogenization data through the public GitHub repository \url{https://github.com/CPShub/sim-data}.
\vspace*{1ex}

\noindent
\textbf{AI declaration.} This manuscript was written independently and linguistically revised with the help of ChatGPT, Grammarly, and DeepL.


\appendix
\numberwithin{equation}{section} 

\section{Notation}\label{app:notation}

Throughout this work, tensor spaces of rank greater than zero are denoted by
\begin{equation}
\cL_n:=\underbrace{\bbR^3\otimes\cdots\otimes\bbR^3}_{n\text{-times}}\,\,\forall n\in\bbN^+\,,
\end{equation}
where $\bbR^3$, $\bbN^+$, and $\otimes$ denote the Euclidean vector space in $\bbR^3$, the set of natural numbers excluding zero, and the dyadic product, respectively. First, second, and fourth order tensors are denoted by $\ba\in\cL_1$, $\bA\in\cL_2$, and $\bbA\in\cL_4$, respectively. 

Tensor compositions and contractions are denoted by $(\bA\cdot\bB)_{ij}=A_{ik}B_{kj}$, $(\bA\cdot\bb)_{i}=A_{ik}b_{k}$, $\ba\cdot\bb=a_ib_i$, $\bA:\bB=A_{ij}B_{ij}$, and $(\bbA:\bA)_{ij}=\bbA_{ijkl}A_{kl}$, respectively, where Einstein's summation convention is applied. 
The tensor cross product operator $\Cross$ is defined as $(\bA\Cross\bB)_{iI}=\mathcal{E}_{ijk}\mathcal{E}_{IJK}A_{jJ}B_{kK}$, where $\mathcal{E}_{IJK}$ is the permutation symbol. 
The transpose and inverse of a second order tensor $\bA$ are denoted by $\bA^T$ and $\bA^{-1}$, while trace, determinant, and cofactor are denoted by $\tr\bA$, $\det\bA$, and $\cof\bA:=\det(\bA)\bA^{-T}$, respectively. Norms of tensors of order one and two are given by $\norm{\ba}=\sqrt{a_ia_i}$ and $\norm{\bA}=\sqrt{A_{ij}A_{ij}}$, respectively.
The  first Fr\'echet derivative of a function $f$ w.r.t.\ $\bA$ is denoted by $\partial_{\bA}f$, the second Fr\'echet derivative w.r.t.\ $\bA$ and $\bB$ is denoted by $\partial^2_{\bA\bB}f$. Partial derivatives are denoted by $\partial_{\bA}f$, total derivatives are denoted by $d_{\bA}f$.
The gradient operator is denoted by $\nabla=\partial_{x_k}(\bullet) \be_k$ and the divergence operator by $\nabla\cdot$. A zero as a subscript next to the nabla symbol, i.e., $\nabla_0$, indicates that the operation is carried out in the reference configuration.

As common in the literature, we do not explicitly distinguish between the algebraic structure of a group and the associated set. For example, $\cG$ can stand for the group given by the tuple $(\cG,\,\cdot)$ or the set $\cG$ of tensors involved therein. We consider in the $\bbR^3$ the general linear group $\GL^+(3):=\big\{\bA \in\allowbreak \;\cL_2\,\rvert\,\allowbreak \det \bA >0\big\}$, the space of symmetric positive definite tensors $\SYM^+(3):=\big\{\bA \in\allowbreak \;\cL_2\,\rvert\,\allowbreak \bA=\bA^T,\,\ba\cdot\bA\cdot\ba>0\,\forall\ba\in\cL_1\backslash{\{\bnull}\}\big\}$, the orthogonal group $\operatorname{O}(3):=\big\{\bA \in\allowbreak \cL_2\;\rvert\allowbreak \;\bA^T\cdot\bA=\bI\big\}$, and the special orthogonal group $\SO(3):=\big\{\bA \in\allowbreak \cL_2\;\rvert\allowbreak \;\bA^T\cdot\bA=\bI,\;\det \bA =1\big\}$, where $\bI\in\cL_2$ denotes the second order identity tensor. We consider the fourth-order unit tensor $\mathbb{I}$ with $\bA=\mathbb{I}:\bA$ for all $\bA$.
To enhance readability, function arguments are omitted throughout this work unless their inclusion is required for clarity.

\newpage

\renewcommand*{\bibfont}{\footnotesize}
\printbibliography
\end{document}